\documentclass[useAMS,usenatbib]{mnras} 
\usepackage{graphicx}
\usepackage{amsfonts} 
\usepackage{natbib}
\usepackage{amssymb, amsmath, amsbsy}
\usepackage[dvips]{color} 
\usepackage{setspace} 

\title[Numerics and Cooling Physics in Cosmological Simulations]{The
Robustness of Cosmological Hydrodynamic Simulation Predictions to
Changes in Numerics and Cooling Physics} \author[S. Huang et al.]{Shuiyao
Huang$^{1}$\thanks{E-mail:shuiyao@astro.umass.edu}, Neal Katz$^{1}$,
Romeel Dav\'e$^{2,3,4}$, Mark Fardal$^{1}$, Juna Kollmeier$^{5}$,
\newauthor Benjamin D. Oppenheimer$^{6}$, Molly S. Peeples$^{7,8}$, Shawn
Roberts$^{1}$, David H. Weinberg$^{9}$, \newauthor Philip F. Hopkins$^{10}$ and Robert Thompson$^{11}$ 
\\$^1$ Astronomy Department, University of Massachusetts, Amherst, MA 01003, USA
\\$^2$ University of the Western Cape, Bellville, Cape Town 7535, South Africa
\\$^3$ South African Astronomical Observatories, Observatory, Cape Town 7925, South Africa 
\\$^4$ Institute for Astronomy, University of Edinburgh, Royal Observatory, Edinburgh EH9, 3HJ, UK 
\\$^{5}$ Observatories of the Carnegie Institution of Washington, 813 Santa Barbara Street, Pasadena, CA, 91101, USA 
\\$^{6}$ CASA, Department of Astrophysical and Planetary Sciences, University of Colorado, Boulder, CO 80309, USA
\\$^{7}$ Space Telescope Science Institute, Baltimore, MD 21218
\\$^{8}$ Johns Hopkins University, Department of Physics \& Astronomy
\\$^{9}$ Astronomy Department and CCAPP, Ohio State University, Columbus, OH 43210, USA
\\$^{10}$ TAPIR, Mailcode 350-17, California Institute of Technology, Pasadena, CA 91125, USA 
\\$^{11}$ Portalarium, 3410 Far West Blvd, Austin, TX, 78731, USA 
}

\begin{document}

\date{Accepted 0000 October 00. Received 0000 October 00; in original
form 0000 October 00}

\pagerange{\pageref{firstpage}--\pageref{lastpage}} \pubyear{0000}

\maketitle

\label{firstpage}


\begin{abstract}
We test and improve the numerical schemes in our smoothed particle hydrodynamics (SPH) code for cosmological simulations, including the pressure-entropy formulation (PESPH), a time-dependent artificial viscosity, a refined timestep criterion, and metal-line cooling that accounts for photoionisation in the presence of a recently refined Haardt \& Madau (2012) model of the ionising background. The PESPH algorithm effectively removes the artificial surface tension present in the traditional SPH formulation, and in our test simulations it produces better qualitative agreement with mesh-code results for Kelvin-Helmholtz instability and cold cloud disruption. Using a set of cosmological simulations, we examine many of the quantities we have studied in previous work. Results for galaxy stellar and HI mass functions, star formation histories, galaxy scaling relations, and statistics of the Ly$\alpha$ forest are robust to the changes in numerics and microphysics. As in our previous simulations, cold gas accretion dominates the growth of high-redshift galaxies and of low mass galaxies at low redshift, and recycling of winds dominates the growth of massive galaxies at low redshift. However, the PESPH simulation removes spurious cold clumps seen in our earlier simulations, and the accretion rate of hot gas increases by up to an order of magnitude at some redshifts. The new numerical model also influences the distribution of metals among gas phases, leading to considerable differences in the statistics of some metal absorption lines, most notably NeVIII.
\end{abstract}

\section{INTRODUCTION} 
Numerical simulations
are indispensable for our understanding of galaxy formation and
evolution. In the $\mathrm{\Lambda CDM}$ model, dark matter haloes over a wide
mass range form and grow through gravitational instability. Star formation
and galaxy assembly take place within these gravitational potentials,
where baryonic processes such as gas accretion, shock heating, and
cooling are essential. Modelling these baryonic processes accurately
is thus crucial. Cosmological simulations with
simplified treatments of baryonic physics such as radiative cooling and photoionisation \citep{katz96a}, star formation \citep{springel03}, chemical
enrichment, and supernova feedback \citep[e.g.,][]{oppenheimer06} have enjoyed many successes
matching observational results on various spatial and time scales
\citep{hernquist96, dave01, governato07, oppenheimer06, oppenheimer08,
dave13}. To make simulations faithfully
represent the true universe, we need not only realistic prescriptions
for the sub-grid physics, but also accurate and stable hydrodynamic
solvers so that the simulated gas thermodynamics accurately converges
to the behaviour of gas in the real physical world.

The two basic methods that evolve hydrodynamical systems are Eulerian
based and Lagrangian based, which differ in how the fluid equations
are discretised. The smoothed particle hydrodynamics (SPH) technique
\citep{gingold77, lucy77, monaghan92} discretises the fluid elements
into particles that carry local thermodynamic quantities that are
evaluated using kernel smoothing. The equations that govern the motion
of SPH particles are derived rigorously from the discretised Lagrangian,
automatically satisfying the continuity equation, and are symmetrised to
guarantee a simultaneous conservation of energy, momentum and entropy
\citep{springel02}. The pseudo-Lagrangian nature of SPH allows it to
probe a large dynamic range in the cosmological context, and makes it
convenient in studying accretion events and outflow models. Also, the
particle based hydro-force calculation enables a straightforward coupling
with many efficient algorithms for the gravity force calculation. However,
SPH techniques have well-known weaknesses including poor shock
resolution, over-damping of weak non-convergent velocity fields, and suppression of fluid
instabilities. Mesh based codes solve the Eulerian hydrodynamical
equations on a grid that divides the simulation volume. A weakness
of grid-based codes is the lack of spatial dynamic range needed for a
representative size of the universe. Adaptive mesh refinement (AMR) codes
\citep{berger89, teyssier02, bryan14} are designed to alleviate this problem,
but still have other problems such as over mixing, not being Galiliean invariant, and poor coupling to gravity solvers \citep{vogelsberger12,  gizmo}. Some more recently developed algorithms, such as the moving mesh code AREPO \citep{springel10a} or the Lagrangian volume method implemented in GIZMO \citep{gizmo}, attempt to take advantage of the best properties of SPH and AMR. Assessing the relative merits of these schemes versus the overall class of SPH methods is a long-term project and one we will not address here.

The two classes of methods can give very different results in both standard
test problems and cosmological simulations \citep{frenk99, hu14, sembolini16a, sembolini16b}, and the disagreements are not alleviated by
simply increasing the resolution \citep{agertz07}. In sub-sonic regimes,
traditional SPH is believed to lead to unphysical results especially
in regions where two fluids of strong density contrast intersect. The
poor behaviour of SPH at fluid interfaces has been attributed to an
erroneous pressure force analogous to a surface tension, which is
caused by multi-valued pressures at contact discontinuities. Many
modifications to traditional SPH have been proposed to alleviate this
problem \citep{abel11, price12, read12}. However, these methods have
certain problems such as increased run time, and a requirement for high
order smoothing kernels that need a large number of neighbouring particles
to keep an equivalent resolution. Furthermore, some methods introduced
additional terms that violate the conservation properties of SPH \citep{peeples10}.

\citet{saitoh13} pointed out that by using an alternative definition of
the SPH volume element, a new set of equations can be derived to eliminate
the surface tension term. Following their work, \citet{hopkins13}
derived a class of alternative SPH equations of motion from the discrete
Lagrangian. His work shows that, without losing general conservation
properties, this pressure-entropy formalism of SPH (referred to below as PESPH) makes significant improvements to the code's performance
at contact discontinuities. His tests show that the improvements are
largely attributed to the optimal choice of hydrodynamical equations
while the assumptions on smoothing kernel and artificial viscosity only
have sub-dominant effects. Further improvements are realised if one also includes artificial conduction. 

Another problematic aspect of SPH is the artificial viscosity. Owing to the entropy conserved nature of the Lagrangian fluid equations, an extra artificial viscosity term has to the added to the momentum equation of SPH to efficiently convert kinetic energy into thermal energy in shock regions. Usually the artificial viscosity needs to be much larger than the physical viscosity to avoid particle penetration between SPH particles in strong shocks. The traditional treatment of artificial viscosity \citep{springel05}, however, usually leads to a fluid that is too viscous, in which small velocity perturbations away from shocks are over damped. Therefore, a modified treatment of artificial viscosity that reduces unwanted dissipation in shock-free regions is motivated. \citet{mm97} (hereafter M\&M97) proposed a time-dependent method that adjusts the viscosity by a pre-factor that depends on the strength of the local convergent flow. \citet{cd10} presented a more complicated method that produces more accurate results in certain test problems, but the behaviour of their method in a cosmological context is still uncertain.

Meanwhile, a long standing problem in current galaxy formation theory is that cosmological simulations without external feedback processes have produced too many stars compared to observations \citep{keres05, keres09a}. Recently, various forms of feedback have been adopted in studies of galaxy formation to suppress star formation. However, as pointed out by several authors \citep{saitoh09, merlin10}, the traditional timestep criterion is inaccurate in handling strong perturbations that arise from feedback prescriptions, leading to violations of energy conservation. Therefore, a SPH code using adaptive timesteps must ensure a prompt response of the system to strong energy perturbations from such feedback. \citet{durier12} proposed a timestep limiter to address this problem for both thermal and kinetic feedback and demonstrated its success in standard tests such as the Sedov blast wave problem. They found that without properly limiting the timesteps errors can be as large as orders of magnitude.

Almost all of the above mentioned problems, solutions and tests have
been performed on idealised problems that may or may not be important
to simulations of galaxy formation. Furthermore, most of the tests are
run at resolutions that are much higher than that usually obtained
in galaxy formation simulations, whether in zoom-in simulations
\citep{katz93, governato07, guedes11, hopkins14} or in cosmological
volumes \citep{vogelsberger14, schaye15, mufasa}. In addition, choices
in feedback algorithms may dominate over differences that result
from the choice of hydrodynamics solver \citep{scannapieco12, schaller15, sembolini16b}.

To study the effects of these numerics in more realistic situations, in
this work we run a series of simulations using the GADGET-3 code, which is
an updated version of the widely used code GADGET-2 \citep{springel05}. It
is a hybrid Tree-PM SPH code that aims at studies of gas dynamics
within a gravitational background. The long range gravitational forces
are evaluated using a particle mesh (PM) method \citep{hockney81}
and the short range forces are evaluated with an oct-tree algorithm
\citep{barnes86}. The SPH in the original GADGET-3 uses the old
density-entropy formalism \citep{springel02}, which manifestly conserves
energy and entropy, but we have replaced the density-entropy SPH
equations (referred to below as DESPH) with the new pressure-entropy SPH equations.

In our previous hydrodynamical cosmological simulations, we compute the
metal-line cooling rate with the assumption of collisional ionisation
equilibrium (CIE). The CIE assumption is, however, not rigorous since
photoionisation reduces the amount of bound electrons and thus affects
the cooling rates. \citet{wiersma09} computed non-CIE cooling rates in the
presence of a radiative background and showed that photoionisation could
suppress cooling in shocked gas by an order of magnitude. Furthermore,
\citet{haardt12} (HM12) published a more up-to-date estimate of the
UV background flux with a more careful assessment of the galaxy and
quasar contributions as a function of time. The effects of these recent
updates in the sub-resolution physical models has yet to be studied in
a cosmological context.

The main goal of this paper is to examine the impact of the new numerical
treatments of hydrodynamics as well as the adopted prescriptions
for the baryonic physics in a full cosmological simulation. We also
perform standard tests to demonstrate the improvements made by the
new SPH formalism. The simplified physics in the standard tests help us to
understand the behaviour of different codes as one changes the numerical
resolution. Since the relevant physics such as
contact discontinuities, fluid instabilities, and mixing are prevalent in
cosmic baryonic processes, such knowledge is important to interpreting
the predictions of the baryons in the realistic yet much more complex
problems of structure formation and evolution. We have published many
predictions using DESPH with the old viscosity, cooling and UV background
\citep{oppenheimer06, finlator08, oppenheimer08, oppenheimer10, dave10, peeples10a, peeples10b, dave11a, dave11b, oppenheimer12, dave13, ford13, kollmeier14, ford16},
and would like to show which of those predictions are robust to these
numerical and physical changes and which have been altered. This retrospective comparison is a necessary prelude to our future work using the new SPH code.

\citet{schaller15} compare a subset of EAGLE cosmological simulations that use traditional SPH and fiducial ANARCHY flavour of SPH, which includes the PESPH formulation, a simplified \citet{cd10} viscosity, artificial conduction and the timestep limiter. They conclude that the numerical improvements included in the ANARCHY SPH do not have significant effect on the properties of most galaxies. However, they use a thermal feedback scheme which is very different from our kinetic feedback scheme. The differences in the feedback model could result in very different gas properties in the universe that are likely to be sensitive to the numerical techniques. Therefore, our work is an independent test of the importance of numerics on the outcomes of cosmological simulations.

The paper is organised as follows. \S 2 describes our improvements to our
numerical algorithms, including the new PESPH formulation, the artificial
viscosity, artificial conduction, the timestep limiter, and the implementation of these changes
into our cosmological code. We perform standard fluid dynamics tests
with our updated code, and present the results in \S 3. We describe our
cosmological tests - the subgrid physics, the simulation parameters,
etc. in \S 4. In \S 5 we compare baryonic statistics such as stellar
mass - halo mass relations, hot baryon fractions, baryonic accretion
histories and the stellar mass functions, which result from our
cosmological simulations, and in \S 6 we focus on comparing the properties
of the intergalactic and circumgalactic medium, including Ly$\alpha$
statistics, metal line absorption lines. In all these comparisons we
focus on the predictions that are the most sensitive to change, in both
the numerics and physics. We present a summary in \S 7. In the Appendix
we present results for those predictions that are not much affected by
changes in the numerics and physics, and some additional idealised tests.

\section{THE NEW HYDRODYNAMICS}
\subsection{A new hydro-solver}
\label{sec:pesph}
A
comprehensive review of the standard SPH formalism can be found in
\citet{springel10b}. The SPH equations of motion are derived from the
Lagrangian form of the fluid equations. Each SPH particle represents a
fluid element of small volume $\Delta V_i$, which defines the size of
the smoothing kernel and connects the thermodynamic quantities, e.g.,
pressure and specific entropy. In traditional DESPH, the volume element is
always assumed to be $m_i/\rho_i$, in which the density of the particle
$\rho_i$ is computed by kernel averaging over its $N_{ngb}$ neighbouring
particles, and $m_i$ is the mass of the particle. \citet{hopkins13} highlights the freedom in the choice of this volume element
without violating the conservation properties. A general way of defining
the volume element can be written as the ratio of a particle-carried
scalar value $x_i$ and its kernel averaged value $y_i$:
\begin{equation}
  \label{eqn:volume}
  \Delta V_i = \frac{x_i}{y_i}
\end{equation}
\begin{equation} y_i \equiv
  \sum_{j=1}^{N_{ngb}} x_iW_{ij}(h_i)
\end{equation} where $W_{ij}(h_i)$
is the smoothing kernel for particle i. \citet{springel02} found that the
SPH equations derived from the discrete particle Lagrangian are able to
conserve energy, momentum and entropy simultaneously if the smoothing
length $h_i$ of each particle satisfies the constraint equation with
a constant $N_{ngb}$ for all particles:

\begin{equation}
  \label{eqn:volume_tilde}
  \Delta \tilde{V} = 
  \frac{4\pi}{3}h_i^3\frac{1}{N_{ngb}}
\end{equation}

The two volume elements, $\Delta V$ and $\Delta \tilde{V}$, defined through equation \ref{eqn:volume} and equation \ref{eqn:volume_tilde}, respectively, do not have to be the same in a simulation. By taking into account the arbitrariness of the volume element choices, \citet{hopkins13} derived the
general form of the equations of motion (EoM): \begin{equation}
m_i\frac{d\mathbf{v_i}}{dt} = -\sum_{j=1}^{N_{ngb}}
x_ix_j\left[\frac{P_i}{y_i^2}f_{ij}\nabla_iW_{ij}(h_i)
+ \frac{P_j}{y_j^2}f_{ji}\nabla_iW_{ij}(h_j)
\right] \end{equation} \begin{equation} f_{ij} \equiv 1 -
\frac{\tilde{x_j}}{x_j}\left(\frac{h_i}{3\tilde{y_i}}\frac{\partial{y_i}}{\partial{h_i}}\right)\left[
1 + \frac{h_i}{3\tilde{y_i}}\frac{\partial{\tilde{y_i}}}{\partial{h_i}}
\right]^{-1} \end{equation}
where $\mathbf{v_i}$ is the velocity
and $P_i$, $P_j$ are the pressure, $\tilde{x_i}$ and $\tilde{y_i}$ are associated with the volume $\Delta \tilde{V_i}$ that is defined through the constraint equation \ref{eqn:volume_tilde}. The equations of motion for
traditional SPH (DESPH) are recovered when the volume elements are
defined as $x_i=\tilde{x_i}=m_i$, and $y_i=\tilde{y_i}=\rho_i$. However,
\citet{hopkins13} suggests an alternative choice, the pressure-entropy
formulation: $x_i = m_iA_i^{1/\gamma}, \tilde{x_i}=1(\Delta
\tilde{V}=1/n_i)$. In this formulation, which we refer to below as
pressure-entropy SPH (PESPH), the quantity that is evaluated from
direct kernel smoothing is now the pressure instead of the density:
\begin{equation} P_i = y_i^{\gamma} = \left[ \sum_{j=1}^{N_{ngb}}
m_jA_j^{1/\gamma}W_{ij}(h_i)\right]^{\gamma} \end{equation} where $A_j$ is
the specific entropy. With these choices the equations of motion become:

\begin{align}
  m_i\frac{\mathbf{v_i}}{dt} = -\sum_{j=1}^{N_{ngb}} & m_j \left[
\left( \frac{A_j}{A_i} \right)^{\frac{1}{\gamma}} + f_i \right] \left(
\frac{A_i}{P_i} \right)^{\frac{2}{\gamma}} P_i\nabla_iW_{ij}(h_i) + \\
                                               & m_j \left[ \left(
                                               \frac{A_i}{A_j}
                                               \right)^{\frac{1}{\gamma}}
                                               + f_j \right] \left(
                                               \frac{A_j}{P_j}
                                               \right)^{\frac{2}{\gamma}}
                                               P_j\nabla_iW_{ij}(h_j)
\end{align}
where,
\begin{equation} f_i = \frac{h_i}{3A_i^{1/\gamma}m_in_i}
+ \frac{\partial{P_i}^{1/\gamma}}{\partial{h_i}}\left[ 1 +
\frac{h_i}{3n_i}\frac{\partial{n_i}}{\partial{h_i}} \right]^{-1}
\end{equation}

The density, now derived from the pressure, resembles an entropy weighted
kernel average rather than a direct smoothing over its neighbouring
particles: \begin{equation} u \equiv \frac{P_i}{(\gamma -1)\rho_i} =
\frac{A_i\rho_i^{\gamma-1}}{\gamma - 1} \end{equation} where $u$ is the
specific energy.

\subsection{Kernel choice}
\label{sec:kernels}
The equations of motion allow for an arbitrary
choice of the smoothing kernel $W_{ij}(r_{ij}, h_i)$. A standard
cubic spline kernel with 32 neighbouring particles ($N_{ngb}=32$) has
been adopted in our previous work, which used the DESPH formulation
\citep[e.g.,][]{oppenheimer06, ford16}.

In recent years, high resolution standard tests demonstrate that
different kernels can have a significant impact on certain problems,
so the choice is not trivial. SPH suffers from an intrinsic $O(h^{-1})$
error in the momentum equation \citep{morris96}. This error can be reduced
by increasing the number of neighbours, but without a proper choice for the
kernel function numerical instabilities grow and degrade the results
\citep{read10, dehnen12}. These authors consistently demonstrated that
a higher-order kernel smoothing over a sufficient number of neighbours
provides an effective solution in suppressing the errors and numerical
instabilities. Following \citet{hopkins13}, we choose the quintic spline
kernel with $N_{ngb} = 128$ as our default, which has an effective resolution
equal to a cubic spline kernel with 34 neighbours. This choice is motivated
by test results from \citet{hongbin05} and \citet{dehnen12}.

\subsection{Artificial viscosity} SPH equations are intrinsically
dissipationless in the sense that the entropy of each particle is
conserved when there is no external heat or cooling source. To capture shocks
in real physical situations, an artificial viscosity is added as a
dissipation that converts the kinetic energy into thermal energy
for gas particles in a converging flow. Traditionally a viscosity
force is added to the momentum equation \citep{springel03}:
\begin{equation} \frac{d\mathbf{v_i}}{dt}|_{visc}
= -\sum_{j=1}^Nm_j\Pi_{ij}\nabla_i\bar{W}_{ij}
\end{equation} where, \begin{equation} \Pi_{ij} =
-\frac{\alpha}{2}\frac{(c_i+c_j-w_{ij})w_{ij}}{\rho_{ij}} \end{equation}
where $c_i$ and $c_j$ are the sound speed of particle i and j, respectively, $\rho_{ij}$ is the mean of their densities, and $w_{ij} = \mathbf{v_{ij}}\cdot\mathbf{r_{ij}}/|\mathbf{r_{ij}}|$
whenever particles are approaching each other
($\mathbf{v_{ij}}\cdot\mathbf{r_{ij}} > 0$), otherwise $w_{ij}$ is
set to 0 making for no viscous force. It is important to always convert
comoving coordinates to physical ones whenever applying any artificial
viscosity scheme.

The parameter $\alpha$ that appears in the above equation regulates the
overall strength of the viscous force. It used to be empirically set to
a constant value ($\alpha = 0.2$ in our previous simulations). However,
the standard artificial viscosity often leads to unnecessary damping of
velocity perturbations in regions where turbulence dominates, because
it only requires velocity convergence on a particle by particle
basis. \citet{mm97} proposed that the $\alpha$ parameter of each
individual particle be allowed to vary depending on the local convergence
of the flow. The $\alpha$ is adapted to evolve through the differential
equation: \begin{equation} \dot{\alpha} = (\alpha_{min}-\alpha_i)/\tau_i
+ S_i \end{equation} The decay time-scale, $\tau_{i}$ is related to the
sound crossing time and the source term $S_i$ is based on the local
divergence of the velocity:
\begin{equation}
\tau_i = h_i/(2lc_i)
\end{equation}

\begin{equation}
S_i = S_0 \times max\{-\nabla\cdot\mathbf{v_i}, 0\}
\end{equation}

In our simulations that adopt the
M\&M viscosity, we set the parameters to $l=3.73, \alpha_{min}=0.1$,
and $S_0 = 2.0$. We also imposed an upper limit of $\alpha_{max} = 2.0$,
preventing the viscosity from becoming too large owing to numerical
noise. We also adopt the \citet{balsara95} switch that aims to suppress
viscosity in shear flows in conjunction with the \citet{mm97} viscosity.

For a cosmological simulation in a co-moving volume, we add the Hubble
flow of $3H(a)$ to the divergence to account for the expansion of
the universe. In practise, when we use the new M\&M viscosity in a
cosmological simulation, the $\alpha$'s of the majority of the gas
particles are kept at the lower limit of 0.1, retaining the turbulent
nature of most of the diffuse gas. When a shock occurs, it is captured
by the convergence check and the $\alpha$'s of shocked particles are
boosted to a higher value compared to the traditional scheme, efficiently
converting the shock energy into thermal energy. In the post-shock
region, the $\alpha$'s decay back to the lower limit on a sound-crossing
time-scale, which avoids further damping of the velocity perturbations.

\citet{cd10} recently provided an improved prescription for artificial
viscosity and also a more accurate estimator for $\nabla \cdot
\mathbf{v}$. In their method, a converging flow is predicted by a shock
indicator that depends on the time derivative of the velocity divergence.
\begin{equation} A_i = \xi_i max\left( -\dot{\nabla} \cdot \mathbf{v_i},
0 \right) \end{equation} Here, $\xi_i$ is a limiter similar to the
Balsara switch, which suppresses dissipation in shear flows:

\begin{equation}
\xi_i =
\frac{\left| 2(1-R_i)^4\nabla\cdot\mathbf{v_i}\right|^2}{\left|
2(1-R_i)^4\nabla\cdot\mathbf{v_i}\right|^2 +
tr(\mathbf{S_i}\cdot\mathbf{S_i^t})}
\end{equation}
where $\mathbf{S_i}$ is a
traceless symmetric matrix defined in \citet{cd10} as an alternative to
$\left|\nabla\times\mathbf{v_i}\right|^2$, and $R_i$ is defined as
\begin{equation}
R_i
\equiv \frac{1}{\rho_i}\sum_j m_j\mathrm{sgn}(\nabla\cdot\mathbf{v_j})W(r_{ij},
h_i)
\end{equation}

The viscosity coefficient $\alpha_{loc}$ is computed for each SPH particle
as
\begin{equation} \alpha_{loc,i} = \alpha_{max}\frac{h_i^2A_i}{h_i^2A_i
    + 0.36c_{ij}^2}
\end{equation}
When $\alpha_{loc;i} > \alpha_i$, such as
in a compressive flow, we set $\alpha_i$ to $\alpha_{loc;i}$ so that the
viscosity is switched on for the particle. Once it leaves the shock, the
coefficient decays exponentially to the minimal value:
\begin{equation}
\dot{\alpha_i} =
(\alpha_{loc,i} - \alpha_i)/\tau_i
\end{equation}

\begin{equation}
\tau_i = \frac{h_i}{lv_{sig,i}}
\end{equation}

We tested these different algorithms for artificial viscosity in standard hydrodynamic tests (e.g. appendix A) and use the \citet{cd10} viscosity as our fiducial choice. In addition to the original \citet{cd10} implementation, in cosmological simulations we only allow
the viscosity coefficient to grow within a convergent flow ($\nabla \cdot \mathbf{v} < 0$), similar to that suggested in \citet{hu14} and
\citet{hopkins14}. Without this additional criterion, we found that
particle noise in our cosmological simulations falsely triggers too much
numerical dissipation in the intergalactic medium.

This new implementation detects shocks in advance and, with a suitable
choice of $\alpha_{max}$, effectively prevents numerical errors at the
shock front from propagating to larger scales \citep{power14}.

\subsection{Artificial conduction}
\label{sec:ac}
Artificial conduction was
traditionally proposed as a solution to the local mixing problem
suffered by DESPH. By smoothing out the entropy gradient at
fluid interfaces it alleviates the numerical surface tension and
thus enhances mixing \citep{price08}. \citet{hu14} points out
that in the pressure-entropy formulation of SPH, even though the
surface tension is eliminated, pressure estimates at the shock
front can be very noisy owing to the sharp entropy jump. Therefore,
we implement artificial conduction into our fiducial simulation
following \citet{read12}.  \begin{equation} \dot{A}_{cond,i} =
\sum_j^{N_{ngb}}\frac{m_j}{\rho_{ij}}\alpha_{cond}v_{sig}L_{ij}\left[A_i -
A_j(\frac{\rho_j}{\rho_i})^{\gamma-1}\right]\hat{r_{ij}}\nabla_iW_{ij}
\end{equation} Here, $\rho_{ij} \equiv 0.5(\rho_i + \rho_j)$, and
$v_{sig} = c_i + c_j - 3\omega_{ij}$ whenever the value is positive
and zero otherwise. $L_{ij} \equiv \left| P_i - P_j\right|/(P_i + P_j)$
is a pressure limiter that prevents falsely triggered pressure waves in
hydrostatic equilibrium. $\alpha_{cond}$ is a free parameter that is chosen to be 0.25 in this paper.

It is worth stressing that the artificial conduction is a pure numerical
treatment that smears out fluid discontinuities. In a cosmological
setting, we must ensure that the artificial conductive effect of cooling
is negligible compared to the physical radiative cooling. See the appendix
for more details.

\subsection{New timestep criteria}
\label{sec:limiter}
The time integration of the equations
of motion in GADGET-3 is discretised into successive transformations on
finite timesteps. For each timestep a kick-drift-kick (KDK) leapfrog
operator \citep{quinn97} is applied to hydrodynamical quantities. To
account for the large dynamic range in cosmological time-scales,
gas particles are allowed to have individual timesteps regulated by
certain criteria that are based on local gas properties. In the past,
people used the Courant condition for this purpose. However, feedback in cosmological simulations can strongly alter  the thermodynamic states of individual
particles. Recent studies \citep{saitoh09, merlin10, durier12} note
that the standard time integration schemes inaccurately handle strong
perturbations, leading to energy conservation violations and unphysical
particle penetration.

To alleviate this problem, \citet{durier12} proposed new constraints on
adapting the timesteps for feedback particles. Particles that undergo
a sudden change in their thermal or kinetic states must be allowed to
change timesteps to ensure that they communicate with the neighbouring
particles in the next timestep. Also, they need to be integrated with a
smaller timestep, so that the hydrodynamical interactions with the other
particles can be more accurately computed in response to the change. The
hydrodynamical acceleration of the particle and the maximum signal
velocity are reevaluated right after the feedback processes, so that
the new timestep is computed based on the updated local thermodynamic
state in response to the feedback.

In GADGET-3, following their work, we implement this timestep limiter in
the following manner. In the neighbourhood of particle i that is directly
subjected to feedback (either thermal feedback or launched as a wind),
each particle j is ensured to be integrated over a timestep that is
shorter than or equal to a constant $f_{step}$ times the timestep of
particle i. Following their tests, we set the constant $f_{step}$ to 4
in our simulations. If a particle is inactive when it is affected by the
feedback, it will be activated at the next timestep to ensure a prompt
response. Moreover, the maximum signal velocity and the acceleration of
this particle are recalculated after the feedback routine to determine
the timestep. In our cosmological runs, this new timestep scheme adds 15\%
to the total CPU time, most of it coming from recalculating accelerations.

\section{STANDARD TESTS}
\subsection{The Kelvin-Helmholtz instability test} The Kelvin-Helmholtz
instability (KHI) arises from the interface of two fluids when there
is a difference in velocities at the interface. KHI tests are
challenging for traditional DESPH codes, which tend to over suppress
fluid mixing. At fluid interfaces, a small perturbation can grow, forming
characteristic vortex features, which finally leads to a mixture of
the fluids. The first phase of the evolution of the perturbation can
be characterised by linear growth with a Kelvin-Helmholtz time-scale of:
\begin{equation}
  \tau_{KH} = \frac{(\rho_1+\rho_2)\lambda}{v\sqrt{\rho_1\rho_2}}
\end{equation}
where
$\rho_1$ and $\rho_2$ are the densities of the two fluids, $\lambda$
is the wavelength of the perturbation, and $v$ is the relative shear
velocity. The further evolution is much more complicated owing to its
turbulent nature and must be studied with numerical methods. However,
traditional SPH codes have been known to have a problem reproducing
the characteristic roll-up features and prevents the fluids from
mixing. \citep{agertz07, read10}. As an example, we study two fluids
initially in pressure equilibrium with a constant density contrast of 2
and opposing velocities. We will demonstrate that the new PESPH formalism
significantly improves the code's behaviour.

Following \citet{hopkins13}, we take the
initial conditions from the Wengen multi-phase test
suite\footnote{http://www.astrosim.net/code/doku.php}. The simulation
is carried out in a periodic box with a size of $256\times 256\times 8$
kpc, with roughly $7.4\times 10^5$ particles equally distributed on a
3D regular Cartesian grid. The particles that represent the two fluids
have a density and temperature ratio of 2 ($\rho_2 = 0.5\rho_1$, $T_2 =
2T_1$), and flow in opposite directions along the y axis with a relative
velocity of $160$ km/s. The values are chosen so that the sound speed
$c_{s,2} \sim 8|v_2|$. A sinusoidal velocity perturbation is applied to
the surface with $\delta v_y = 8$ km/s and a wavelength of $\lambda =
128$kpc. One of the simulations also applies artificial conduction with
a coefficient of $\alpha_{cond} = 0.25$.


\begin{figure} 
\centering
\includegraphics[width=0.96\columnwidth]{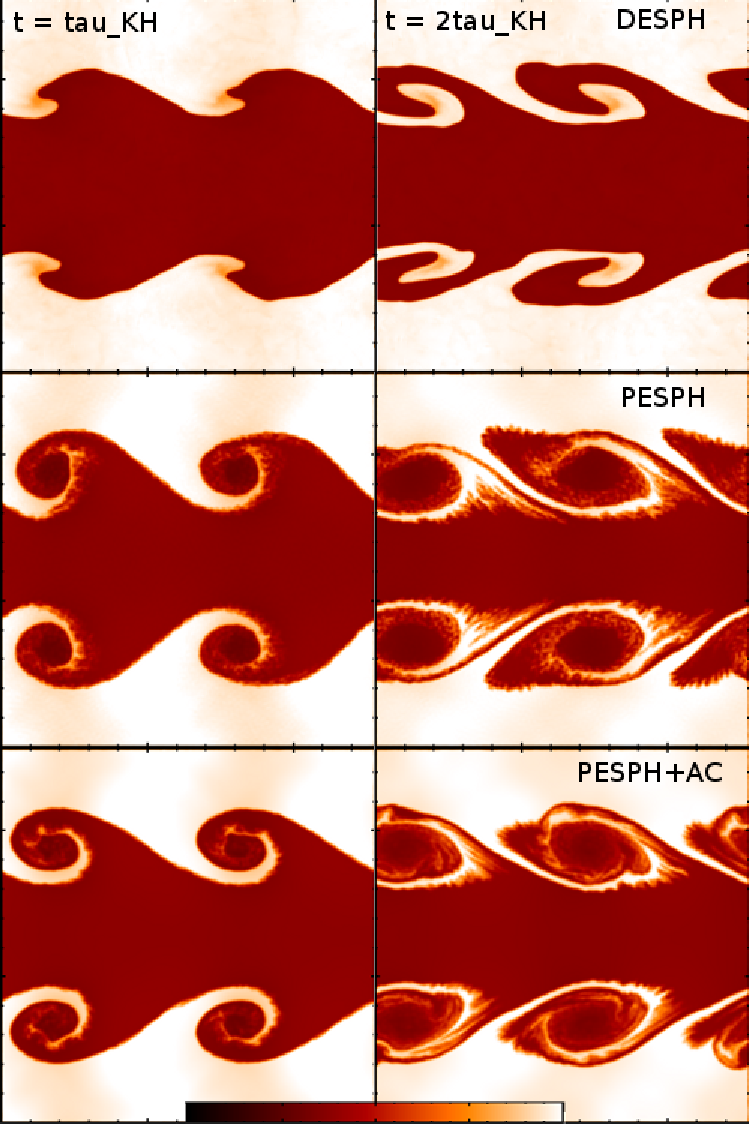}
\caption{A density map of a Kelvin-Helmholtz test at time $t=\tau_{KH}$
(\textit{left}) and $t=2\tau_{KH}$ (\text{right}). \textit{Top}
panels show the results from DESPH using a cubic spline kernel over
$N_{ngb}=32$ neighbours, representing the hydro-solver used in our published
cosmological runs. The \textit{middle} panels show the new PESPH with an
optimal $N_{ngb}=128$ quintic kernel, and the \textit{bottom} panels use
the time-dependent artificial viscosity following \citet{cd10} and artificial conduction, while the
other runs use a constant parameter for the artificial viscosity. All runs
start with an initial condition in which the relative velocity between
two fluids is $160\ \mathrm{km/s}$ and the velocity perturbation is $8\
\mathrm{km/s}$.} \label{fig:khtest} 
\end{figure}

The top and middle panels of Figure \ref{fig:khtest} compare the behaviour
of DESPH and PESPH at $\tau_{KH}$ and $2\tau_{KH}$. The improvements of
PESPH are significant. At the Kelvin-Helmholtz time-scale $\tau_{KH}$,
the characteristic wave-like feature of KHI in the PESPH simulation
have grown to a scale consistent with mesh based results \citep{read10,
murante11} and also the predictions of linear perturbation theory
\citep{agertz07}. The PESPH simulation also shows efficient mixing of the
fluid along with the growth of curled structures. In the DESPH run,
however, the instability barely grows at $\tau_{KH}$, and the mixing is
hardly seen even at later times (not plotted).

In the bottom panel of Figure \ref{fig:khtest} we rerun the PESPH
simulation but switch the prescription of artificial viscosity to
the time-dependent particle-by-particle treatment of \citet{cd10},
and also apply artificial conduction. The new viscosity is known
to significantly reduce unnecessary dissipation owing to velocity
noise from turbulent regions and yet still maintain the ability of
shock capturing. Since the physical conditions in KHI tests are mostly
shock-free, the new viscosity only shows some minor effects at the contact
surfaces at $2\tau_{KH}$. The differences, however, are much smaller than
those between SPH formulations.

\subsection{The blob test} Another classic numerical test involves putting
a spherical gas cloud of uniform density into a fast moving hot medium,
which is in pressure equilibrium with the cloud. It mimics the realistic
situation of a shocked wind passing through cold dense gas, and it
involves important physical processes such as ram-pressure stripping,
fluid instabilities, and mixing. Again we take the initial condition from
the Wengen test suite. The
simulation is performed within a periodic tube with dimensions $x,
y, z = 2000, 2000, 6000$ kpc; the initial density of the cloud is 10
times that of the surrounding medium. The relative velocity of the two
phases of gas is characterised by a Mach number of 2.7.  

\begin{figure}
\centering 
\includegraphics[width=0.96\columnwidth]{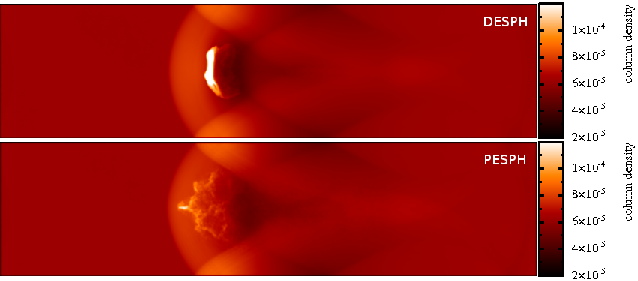}
\caption{ A density map of the blob tests at $\sim \tau_{KH}$ from
the DESPH (\textit{upper}) and PESPH (\textit{lower}) methods.  }
\label{fig:blobtest} 
\end{figure}

\begin{figure} 
\centering
\includegraphics[width=0.96\columnwidth]{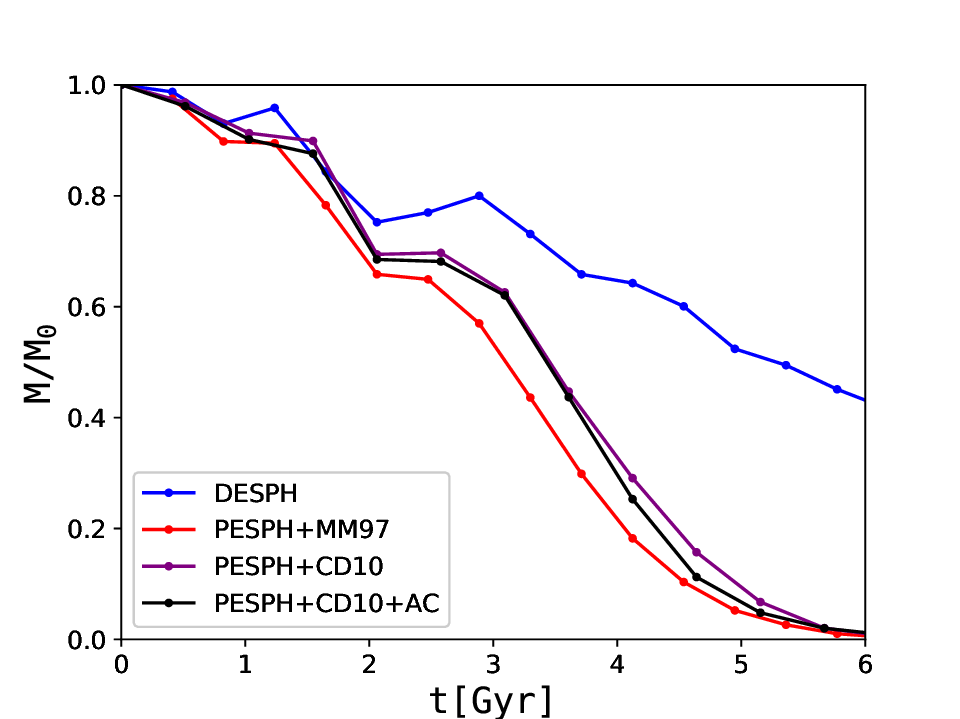} \caption{
The evolution of cloud mass. The different lines correspond to simulations
that use different numerical algorithms as labelled. See text for a full
description of the different models. The cloud quickly loses mass in all
the PESPH simulations owing to enhanced mixing between cold and hot gas.
} \label{fig:mloss} 
\end{figure}

Figure \ref{fig:blobtest} shows the morphology of the gas cloud at
around $\tau_{KH}$. Though the cloud in the DESPH run is deformed owing
to the impact of the wind shock, the cloud particles still stick together,
hardly mixing with the surrounding gas - a behaviour similar to that shown
in the KH tests. In contrast, the clouds simulated using 
PESPH have clearly dissembled
and dissipated into the hot medium. The predicted morphology agrees with
mesh based results \citep[e.g.,][]{agertz07}.

Figure \ref{fig:mloss} shows the mass of the cloud as a function
of time. Following \citet{agertz07}, we define cloud as all those
particles that have a density over 0.64 of its original density ($\rho >0.64\rho_c(t=0)$) and a temperature that is below 0.9 of the ambient
temperature ($T < 0.9T_{amb}$). The blue line uses the DESPH formulation
with a constant viscosity and without artificial conduction. All the
other lines use the PESPH formulation. The red line (PESPH-MM97) uses
the \citet{mm97} viscosity. Both the purple (PESPH-CD10) and the black
line (PESPH-CD10-AC) use the \citet{cd10} viscosity, but the black line
also uses artificial conduction. The evolution of the cloud mass is
similar in all the PESPH simulations and is similar to our results. The clouds in these simulations
lose mass quickly and are mostly destroyed after $6\ \mathrm{Gyrs}$,
when we stop the simulations. The cloud in the DESPH simulation, however,
loses mass at a slower rate and retains more than 40\% of its mass at
the end of the simulation.

In summary, with the new pressure-entropy formalism of the equations
of motion, the performance of our SPH code in both the KHI tests and
the blob test agrees with those from the mesh-based codes \citep[e.g.][]{agertz07, sijacki12}, and agrees
with the improvements seen in \citet{hopkins13} and \citet{gizmo},
who also showed it improved many other tests. These results demonstrate
the successes of PESPH in effectively removing the artificial surface
tension at fluid interfaces that leads to over suppression of fluid
instabilities and mixing, which has been a long-standing problem with the
DESPH formalism. Our results also show that our choices of artificial
viscosity and artificial conduction only has a sub-dominant effect in
these standard tests (see Appendix A for details).


\section{COSMOLOGICAL SIMULATIONS} The standard tests above have shown
that the new PESPH formalism leads to significant improvements in
correctly resolving fluid mixing at contact discontinuities. Now we ask
the question of whether these improvements would significantly change the
results from realistic cosmological simulations. Most of the
idealised tests are conducted at resolutions that are far higher than
those typically obtained in galaxy formation simulations. At realistic
resolutions the differences could be much less.  Furthermore, the dynamical
processes in cosmological simulations involve 
complicated interactions between multiphase
baryons, e.g., cold gas accretion through cosmic filaments, hot halo gas
formed by shock heating, cooling flows from the hot gas, and galactic winds. The hydrodynamical
instabilities and gas mixing could change both the accretion rate and
average cooling efficiency within the halo and possibly alter the entropy
structure as well as the star formation history of the halo. For example,
in previous SPH simulations, e.g., \citet{keres09a}, sub-resolution
clumps of cold gas are shown to orbit within the hot haloes and rain
down upon the central galaxy. This ``cold drizzle'', which is thought to
be a numerical artefact owing to the inefficiency of standard SPH in
multi-phase mixing and stripping \citep{keres12, nelson13}, could artificially enhance cool gas
accretion and the star formation rate. Therefore, the poor numerical
behaviour of the traditional SPH for these important processes adds
uncertainties to the interpretation of the simulation results. However,
given all the other important non-linear processes, such effects could
be sub-dominant. Therefore, it is essential to examine how the changes
introduced by PESPH may affect cosmological conclusions drawn from our previous simulations.

\subsection{Numerical models}
For cosmological simulations, the GADGET-3
code calculates the gravitational forces between all particles using
a tree-particle-mesh algorithm, which allows the code to probe a large
dynamic range efficiently. The dynamics of the gas particles is further
determined by the SPH algorithm. Physical processes such as cooling and
heating, star formation, feedback and metal enrichment play crucial roles
in galaxy formation and evolution. However, to understand these processes,
we must add sub-grid models on top of the hydrodynamical equations,
because they occur on scales much below the spatial resolution of the
simulation. Some of the sub-grid models have led to successes in matching
observed data, and have become routinely incorporated in cosmological
simulations. The specific models that we use have been described in
detail in \citet{oppenheimer06} and \citet{oppenheimer08}. Here we give
a brief summary.

Apart from the dynamical heating processes owing to adiabatic compression and viscosity, an additional heat source implemented in our code is photoionisation whose rate depends on the UV background at different redshifts. To compute the radiative cooling, we tabulate the cooling rate as a function of discrete values of density and temperature, assuming ionisation equilibrium and primordial composition and interpolate. We allow the thermal energy of each particle to change with a cooling rate obtained from this look up table based on its thermodynamic properties. The look up table is updated to account for changes in the radiation background with redshift. When the particle is metal enriched, an additional metal cooling term is added to the total cooling rate.

In our previous work, the cooling rate is tabulated as a function of
metallicity and density from the collisional ionisation equilibrium models
of \citet{sutherland93} in the presence of the \citet{haardt01} (HM01)
ionising background. \citet{wiersma09} computed radiative cooling rates
from 11 elements under the CMB and HM01 background. Subsequently, 
\citet{haardt12} updated their estimate of the UV background by including
several new components to their radiative transfer code CUBA. Motivated
by these recent works, we have adopted the 
\citet{wiersma09} non-CIE cooling rates computed in the presence of the
HM12 background. One of our main goals here is to investigate how this
simulation, named as PESPH-HM12, differs from simulations that employ
the hydrodynamics and cooling model that we used previously.

Our star formation model is adopted from
\citet{springel03}. In this sub-grid model, a gas particle is treated as
a two-phase particle, i.e., it becomes an ISM particle, when the overall
density of the particle reaches a certain threshold. Based on the model
of \citet{mckee77}, we treat the gas particle as if it contains many
cold clouds in pressure equilibrium with the surrounding warm ionised
intermedium gas. The thermodynamic properties of the two phases are
calculated separately following on analytical treatment of evaporation
and condensation. These ISM particles are the sites where stars can
form. The star formation rate is proportional to the square of the
cold phase density and is calculated on a particle-by-particle basis,
with the star formation time-scale fixed to match the observed Kennicutt
law \citep{kennicutt98}. Collisionless particles representing groups of
stars are allowed to form at each timestep, with a probability determined
from the star formation rate. The star particle is either spawned from an
ISM particle, taking away a fraction of its mass, or entirely converted
from the gas particle, depending on how much mass is left in the ISM
particle. The feedback from type II supernovae is added back to the hot
phase assuming an instantaneous recycling approximation.

The metal enrichment model is an updated
version of \citet{oppenheimer08}, in which three main sources of metals,
type II SNe, type Ia SNe and AGB stars, are considered. The simulation now
tracks the metallicity of four species C, O, Si, Fe separately, to account
for the different enrichment effects of the alternative sources. At
each timestep, each ISM particle is self-enriched by type II SNe, which
occurs at a rate proportional to the star formation rate, assuming an
IMF. For the \citet{chabrier03} IMF we adopt in these simulations, we
assume that stars with an initial mass larger than $10M_\odot$ end as
a type II SNe, which gives a mass fraction of 0.18 immediately recycled
upon star formation. The feedback from type Ia SNe and AGB stars returns
metals and mass to the nearest gas particles on a delayed time-scale.

Previous simulations \citep{keres05, keres09b} have shown
that simulations that implement the above ISM and star formation models
produce a global SFR that is too high compared to observations. This
motivates some form of feedback that suppresses either gas accretion or
star formation. A full description of these models is beyond the scope
of this paper but we will summarise the important points. We employ the
hybrid energy/momentum-driven wind model (\textit{ezw} model) \citep{dave13}, which
is a slightly modified version of the momentum-conserving wind model (\textit{vzw}
model) used in our previous work \citep{oppenheimer08, dave10, dave11a,
dave11b}, as our favoured wind model. The \textit{vzw} model and \textit{ezw} model have made
predictions that match a range of observations, including IGM enrichment
at high redshift observed through $C_{IV}$ systems \citep{oppenheimer06,
oppenheimer08}, high redshift absorption systems \citep{oppenheimer09},
mass metallicity relations \citep{finlator08}, and the galactic stellar and HI
mass functions at $z=0$ \citep{dave13}.

We assume in our fiducial outflow model that the outflow rate is related
to the SFR by a mass loading factor $\eta$:
\begin{equation}
  \dot{M}_{wind} = \eta \times SFR
  \end{equation}
The numerical implementation of outflows is analogous to that of
star formation. Each ISM gas particle is a candidate for launching a wind,
the probability of which is $\eta$ times the probability assigned for
star formation, and is probabilistically determined for each particle at
each timestep. Once it is launched, a velocity boost of $v_w$ is added to
the particle in the direction of $\mathbf{v_i}\times\mathbf{a_i}$, where
$\mathbf{v_i}$ and $\mathbf{a_i}$ are the velocity and acceleration of
the particle before launch, as outflows are often seen perpendicular
to the disc where interactions with the cold dense ISM are minimised. All
hydrodynamical interactions relating to the particle are turned off for
$1.95\times10^{10}/v_w$ years or until the particle has reached a density
that is below 10\% of the critical density for star formation. Since
the resolution in the ISM region is insufficient to correctly model hydrodynamical interactions \citep{dallavecchia08}, this decoupling
from hydrodynamical forces allows galactic winds to develop and avoids
calculating numerically inaccurate interactions.

The free parameters $\eta$ and $v_w$ are crucial to the wind models. The
scaling for momentum conserving winds are motivated from \citet{murray05},
which suggested a wind speed that scaled with the galaxy velocity dispersion:

\[ \eta=\left\{ \begin{array}{ll}
  \frac{\sigma_0}{\sigma}\frac{75\ \mathrm{km/s}}{\sigma} (\sigma
  \leqslant 75\ \mathrm{km/s})\\ \frac{\sigma_0}{\sigma} (\sigma \geqslant
  75\ \mathrm{km/s})
\end{array} \right.
 \]

 \begin{equation}
   v_w = 4.29\sigma\sqrt{f_L - 1} + 2.9\sigma
 \end{equation}
where $f_L$, depending on the metallicity, is the luminosity factor in
units of the galactic Eddington luminosity, constrained by observations
\citep{rupke05}, and $\sigma_0=150$km/s is a normalisation factor that
is adjusted to match high-redshift IGM enrichment \citep{oppenheimer08}.

The lower limit $\sigma_{vzw} = 75$km/s in the above equation
distinguishes between momentum-driven scalings and energy-driven
scalings. The momentum-driven mass-loading factor, which scales with
$\sigma^{-1}$, applies to relatively large systems where outflows are
driven primarily by the momentum flux from young stars and supernovae
while the thermal energy from supernova is dissipated too quickly to
become dynamically important. However, in dwarf galaxies with a $\sigma$
below this limit, we assume that energy feedback from supernovae starts
to dominate, based on analytical and numerical models by \citet{murray10}
and \citet{hopkins12}. In this energy conserving regime, we assume $\eta
\propto \sigma^{-2}$. As \citet{dave13} show, this hybrid scaling leads
to better agreement with the low mass stellar mass function.

The velocity dispersion $\sigma$ is determined on-the-fly. We identify
galaxies using a friends-of-friends algorithm that binds particles
to their closest neighbours if they are within a linking length of 0.04 times the typical separation between dark matter particles. This linking length applies to all classes of particles, including dark matter, so it incorporates the inner regions of the dark matter halo. The velocity dispersion is evaluated
from the total mass of the galaxy $M_{gal}$: \begin{equation} \sigma =
200\left(\frac{M_{gal}}{5\times10^{12}M_\odot}\frac{H(z)}{H_0}\right)^{1/3}
\mathrm{km s^{-1}} \end{equation} The velocity dispersion could also be
computed for each galaxy explicitly, but uncertainties arise owing to poor
resolution particularly in the inner regions of each galaxy. Moreover, numerical noise would in some cases yield a rather spurious $\sigma$ that
would lead to unphysical results and when calculating $\sigma$ for satellite
galaxies directly it is almost impossible to remove background material
that belongs to the larger central galaxy. Therefore, we use the above empirical
relation given that any errors that arise from using this relation are
sub-dominant to the uncertainties that come from our assumptions of the
wind model itself.

The FoF algorithm is slightly different from what we used before. Since
$\sigma$ and thus the mass of groups is crucial to our feedback model,
we empirically adjust the linking length in the FoF algorithm so that the
mass of groups found using our old and new algorithms are on average the
same, so that the scaling relations that we adopt for our wind prescription still hold.

\subsection{Simulation setup} To separate the effects of changing
numerical and physical details, we have run two classes of simulations:
\textit{nw} simulations that have no stellar feedback, i.e. no galactic winds;
and \textit{ezw} simulations that employ our fiducial momentum-driven wind
model. Each of the two classes contain simulations that use different
numerical algorithms. We have run two \textit{nw}
simulations in which no feedback is added: one (DESPH-nw) with the
original numerical algorithms and the other one (PESPH-nw),
which uses all the improved schemes. For the \textit{ezw} model, we have run the
following simulations: 1. The DESPH simulation, representative of the
simulations we have used in previous papers, uses the traditional
density-entropy formulation from \citet{springel03}, a cubic spline
kernel with $N_{ngb}=32$, and an artificial viscosity scheme with
a constant $\alpha=1.0$ and the Balsara switch. 2. The PESPH simulation uses the pressure-entropy formulation derived
by \citet{hopkins13}, a quintic kernel of $N_{ngb}=128$ and the M\&M
viscosity. The PESPH simulation also employs the timestep limiter from
\citet{durier12}. In the PESPH simulation, we use the entropy weighted
density only to evolve the hydrodynamics. For other density dependent
physical processes such as cooling and star formation, we still use
the unweighted density obtained from kernel smoothing (see below). 
3. The PESPH-HM12 simulation is based on the numerical algorithms of the PESPH simulation, but it applies
non-CIE metal cooling models combined with the HM12 ionisation background,
instead of the CIE cooling model and HM01 background used in the other
simulations. We also perform another simulation identical to PESPH-HM12
except using the old artificial viscosity to discuss its effects
separately. 4. Our fiducial simulation, PESPH-HM12-AC, combines all the
improvements from the other simulations and also employs the \citet{cd10}
viscosity and artificial conduction. See Table \ref{tab:simulations}
for a summary.

\begin{table*}
\centering 
\caption{Simulations and their adopted numerical and physical models}
\label{tab:simulations}
\begin{minipage}{160mm}
\begin{tabular}{@{}lcccccccc} \hline
Simulation & Formulation & Kernel
 & Viscosity & Limiter\footnote{The timestep limiter proposed in \citet{durier12}. See section \ref{sec:limiter} for details.} & W09\footnote{The \citet{wiersma09} non-CIE metal cooling model.} & UVBKG\footnote{The ionising background adopted in the simulation.} & AC\footnote{Artificial conduction. See section \ref{sec:ac} for details.} & Feedback\\
\hline
DESPH-nw & DESPH & CS-32\footnote{Cubic spline kernel using 32 neighbours. See section \ref{sec:kernels} for details.} & $\alpha=1.0$ & No & No & HM01 & No & \textit{nw}\footnote{No feedback.}\\
PESPH-nw & PESPH & QS-128\footnote{quintic spline kernel using 128 neighbours.} & CD10 & Yes & Yes & HM12 & Yes & \textit{nw}\\
DESPH & DESPH & CS-32 & $\alpha=1.0$ & No & No & HM01 & No & \textit{ezw}\footnote{Kinetic feedback scheme that uses the momentum-driven wind model}\\
PESPH & PESPH & QS-128 & M\&M97 & Yes & No & HM01 & No & \textit{ezw}\\
PESPH-HM12 & PESPH & QS-128 & M\&M97 & Yes & Yes & HM12 & No & \textit{ezw}\\
PESPH-HM12-AC & PESPH & QS-128 & CD10 & Yes & Yes & HM12 & Yes & \textit{ezw}\\
\hline \end{tabular}
\end{minipage}
\end{table*}

Our primary goal in this paper is to study how the new physical
prescriptions in cooling and the improved numerical schemes affect
our predictions from the same initial conditions. Therefore we focus
on comparing the fiducial simulation PESPH-HM12-AC and our original
version of GADGET3 - DESPH. The results from the other simulations provide
information on how specific changes to the code affect certain changes in
the simulated universe. For example, as shown later, PESPH when compared
to DESPH demonstrates the efficiency of the new SPH formulation and a
time-dependent viscosity in removing the pseudo dense gas clumps within
galactic haloes; the PESPH-HM12 shows the effects of the new cooling
model and the background.

In any cosmological simulation that uses the PESPH formulation, there are two ways of defining the density of an SPH particle (see section \ref{sec:pesph}). It could be evaluated either by kernel weighted average as in traditional DESPH, or by entropy weighted average as in PESPH. Here we always use the traditional volume-derived density for any of the sub-grid models such as cooling and star formation, and also for all the post analysis on the simulation data. 
This is to avoid
any bias towards high entropy particles that could lead to multi-valued
densities in neighbouring particles.
The entropy weighted density is only applied while solving the hydrodynamics. \citet{oppenheimer18} points out that this density choice could lead to considerable differences in some predictions (see their appendix D).

We assume a $\Lambda$CDM cosmology with parameters $\Omega_m = 0.30,
\Omega_\Lambda = 0.70, h = 0.7, \Omega_b = 0.045, n=0.96, \sigma_8 = 0.8$
for all our simulations, as a compromise between WMAP7 \citep{hinshaw09}
and the Planck \citep{planck13} results. We performed simulations in a
random periodic box of $50 h^{-1}$Mpc on each side that contains $288^3$
gas and $288^3$ dark particles initially. The initial mass of each gas
particle and dark particle are therefore $m_{gas}=9.3\times 10^7M_\odot$
and $m_{dark}=5.3 \times 10^8M_\odot$. All simulations start at $z\sim100$
and evolve down to $z=0$.

\begin{figure} 
\centering
\includegraphics[width=0.96\columnwidth]{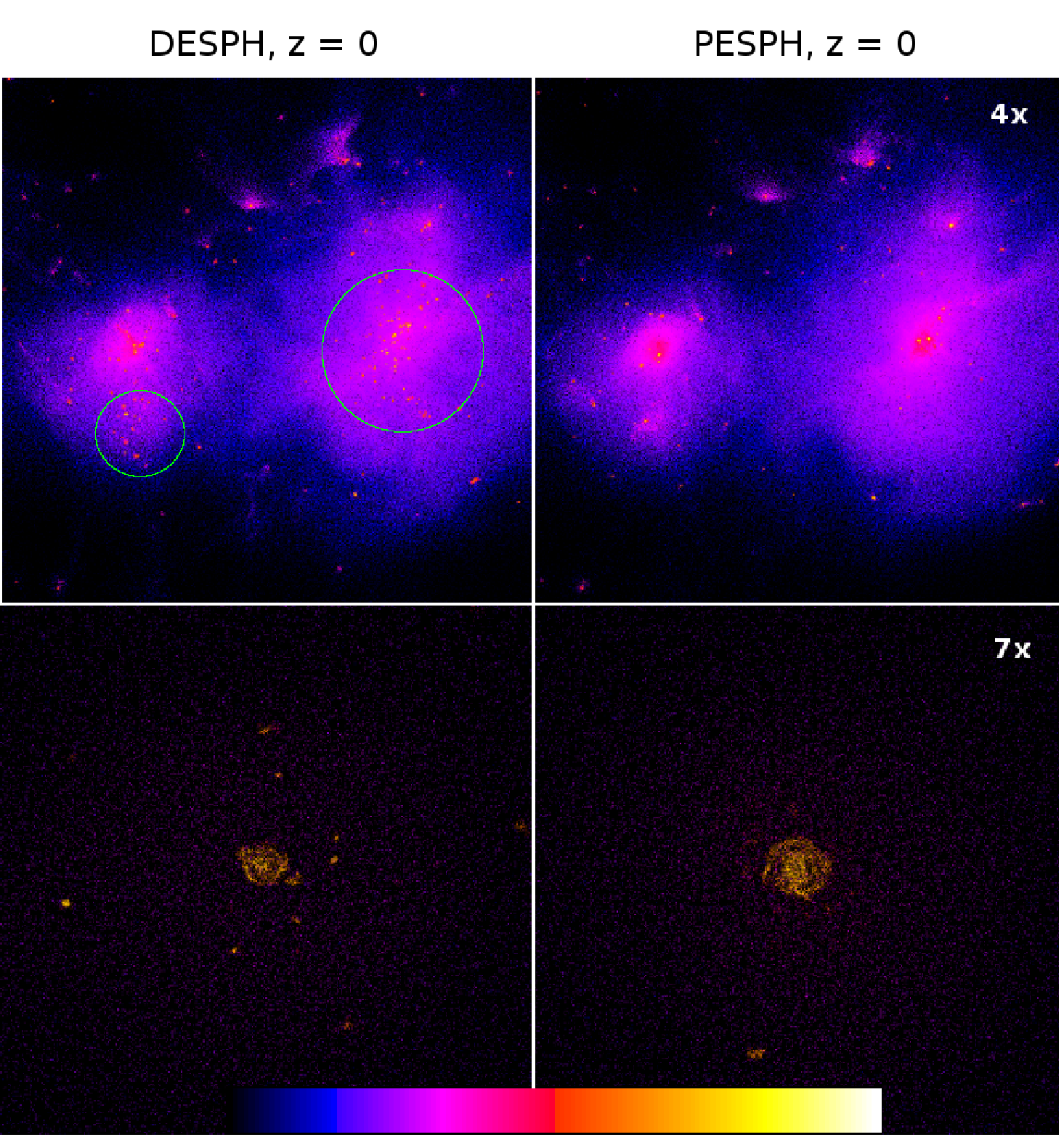}
\caption{Two projected regions are extracted at $z=0$
from the DESPH (\textit{left panels}) and PESPH (\textit{right panels}) cosmological simulations to illustrate the impact
of the new PESPH algorithm. Each panel shows a
density map with a large dynamic range ($\rho/<\rho>=1{\rm\ to\ }10^{10}$). The
simulations are performed within a periodic box with $50\ h^{-1}$ Mpc
on each side. The labels "4x" and "7x" in the upper right corners of
each panel indicate the power of magnification over the original box,
corresponding to a size of $\sim 3.1 h^{-1}$ Mpc and $\sim 0.4 h^{-1}$ Mpc
on both axes, respectively. Green circles highlight the differences
between the snapshots compared.  } \label{fig:snapshots} 
\end{figure}

Figure \ref{fig:snapshots} compares the morphologies of simulated
systems in two regions at $z=0$. The DESPH simulation has
many distinct dense gas clumps orbiting within the diffuse hot
gas halo. These sub-resolution clumps are especially prevalent in the
massive haloes as shown in the upper left panel, where we plot the most
massive halo from the simulation. This feature has also been noticed in the
non-feedback simulations of \citet{keres09a, keres09b}, who attribute
this ``cold drizzle'' to numerical artefacts that would enhance cold mode
accretion within massive haloes. In the PESPH simulation, these
clumps break up and mix into the ambient gas more easily owing
to enhanced fluid instabilities and more efficient mixing between
fluid interfaces, and are thus effectively removed. Therefore, the PESPH formulation alone is sufficient for removing the cold blobs present in the traditional DESPH simulations, consistent with the findings of \citet{hu14}. We also note that,
in the bottom panels, PESPH produces a more extended gas disc in the
centre. \citet{vogelsberger12} and \citet{torrey12} also found that galaxies in their moving
mesh code AREPO are generally more extended and disc-like than those 
in DESPH simulations. However, whether there is a general trend on the
size and morphology of individual galaxies in the PESPH simulation is
beyond the scope of this paper owing to the limited spatial resolution
of the simulations presented here.

\section{Cosmic Baryons}
\subsection{The stellar mass -
halo mass relation} 
Cosmological simulations without any feedback processes have been known
to over-produce the stellar content of the universe \citep{dave01b,
springel03}. Simulations that only consider kinetic feedback from
young stars and supernovae are able to reproduce the z=0 galactic stellar
mass function (GSMF) below $M_* < 10^{10.5}M_\odot$, yet still produce
too many massive galaxies \citep{oppenheimer10}. The cold drizzle,
i.e., dense sub-resolution clumps that do not mix efficiently with hot
surroundings owing to numerical effects, could be accreted by the central
galaxies on a short time-scale and enhance their star formation rate. In
this section we examine to what extent the new numerical schemes affect
the stellar content of the universe, and also the properties of the
simulated galaxies.

To study the statistics of simulated galaxies and haloes, we use SKID
(Spline Kernel Interpolative Denmax) to identify bound groups of star and
ISM particles \citep{keres05, oppenheimer10} as simulated galaxies. We
will focus on those SKID groups that are above the mass limit of $M_* >
6.0\times 10^9M_\odot$, equivalent to a total mass of 64 original gas
particles. This choice is motivated from extensive convergence tests
\citep{finlator06}. Furthermore, we identify haloes using a spherical
overdensity (SO) algorithm \citep{kitayama96, keres05}. In SO, we
start from the most bound particle within each SKID galaxy and search
for all particles within a spherical density contour within which the mean interior density equals the virial density. Then we join haloes to their larger
companions if the centre of the smaller halo falls within the virial
radius of its larger companion. In our simulations, we typically identify
$\sim10^3$ galaxies at $z=4$ and $\sim10^4$ galaxies at $z=0$.

\begin{figure} 
\centering
\includegraphics[width=0.96\columnwidth]{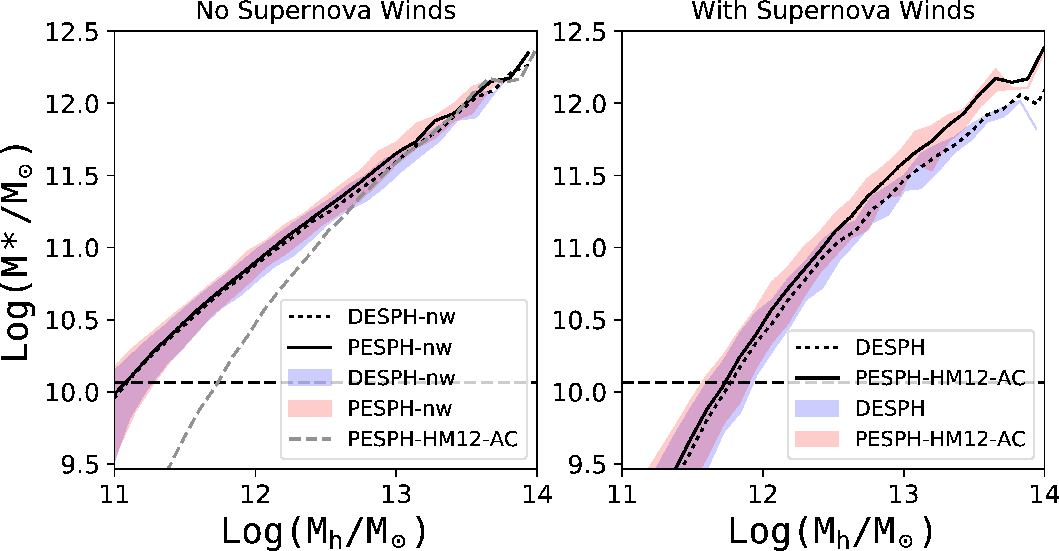} \caption{The
stellar mass - halo mass relations from simulations without supernova
feedback (left panel) and with supernova feedback (right panel). In
each panel, the dotted line and the blue shaded region are from DESPH
simulations, and the solid line and the red shaded region are from PESPH
simulations. The shaded region encloses 95\% of all galaxies in the corresponding mass bin. In the left panel, we reproduce the relation from our
fiducial simulation as a grey dashed line for comparison. The black
dashed horizontal line in each panel indicates the stellar mass resolution
limit for galaxies.} \label{fig:smhm} 
\end{figure}

For each halo found by SO, we derive its virial mass as well as the
mass of its central galaxy. The stellar mass and the halo mass follows
a very tight trend \citep{yang12, moster13} that indicates how efficiently
stars form in haloes of different masses at different times. In Figure
\ref{fig:smhm} we examine how this stellar mass - halo mass relation
(SMHM) is affected by the different numerical schemes.

The SMHMs from simulations that
do not have any feedback (DESPH-nw and PESPH-nw) are distinct from those
that employ the \textit{ezw} momentum-driven wind model. The galaxies in the
\textit{nw} simulations are much more massive in nearly all the haloes and especially so
in less massive ones. Between these two \textit{nw} simulations, however,
the differences are quite small, with PESPH-nw producing only slightly
more stars in the most massive haloes. Compared to the \textit{nw}
simulations, the \textit{ezw} wind model more efficiently suppresses
star formation in lower mass haloes but has hardly any effect on the most
massive ones. This trend with halo mass can be explained by differential
recycling \citep{oppenheimer10}, i.e., the recycling time-scales for
ejected wind particles in low mass haloes are much longer than those
in massive haloes, where the wind particles quickly fall back to the
centre of the potential, resembling a galactic fountain. Like in the
\textit{nw} simulations, the SMHMs in the \textit{ezw} simulations are very similar,
though the differences are more prominent. The median from PESPH-HM12-AC
is less than 0.1 dex higher than DESPH across most of the resolved mass
range. Only in the most massive haloes are the differences larger, where the
galaxies in the PESPH-HM12-AC simulation are around 30\% more massive than their
counterparts with DESPH. In the latter sections we will show that this
owes to an enhanced baryonic accretion rate in the PESPH
simulations.

\subsection{Hot gas fractions} 
The new hydrodynamic schemes have the potential to
alter the properties of the hot gas within galaxy groups, as the PESPH
formulation could enhance the mixing of cold gas into hot corona gas and the new viscosity scheme captures shocks more accurately. Since the
X-ray emitting gas in groups is mostly heated by shocks, its properties
could be sensitive to the numerics.

\begin{figure} 
\centering
\includegraphics[width=1.10\columnwidth]{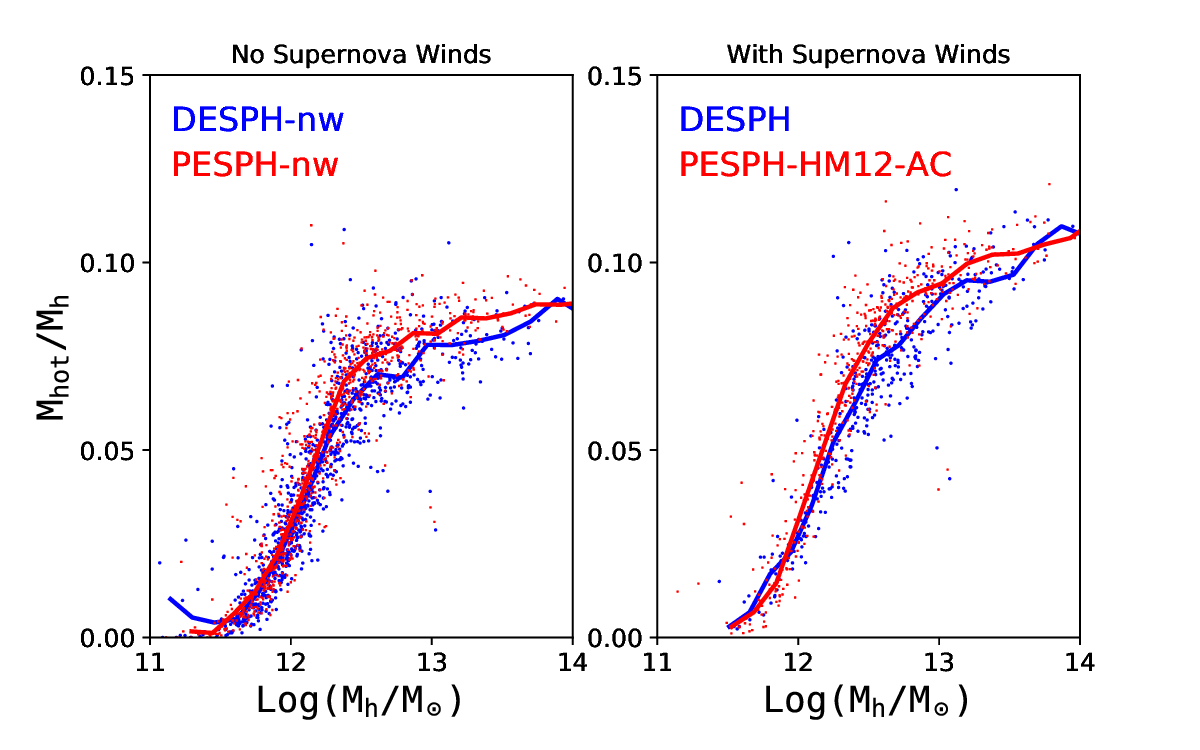} \caption{The
mass fractions of hot gas ($T > 5\times10^5 K$) in simulated haloes as a
function of their
halo mass. Haloes from the DESPH simulations are plotted in blue and
those from the fiducial PESPH simulations are plotted in red. The solid lines show the running medians. In the
left panel we compare the two \textit{nw} simulations and in the right
panel we compare the two \textit{ezw} simulations.} \label{fig:fhot}
\end{figure}

In Figure \ref{fig:fhot} we compare the mass fraction of hot gas ($T >
5\times10^5 K$) in individual haloes. In general, the hot gas fraction
increases with the halo mass because of the deeper potential and higher
virial temperature. The relations in both \textit{nw} simulations
agree surprisingly well. The agreement between the two \textit{ezw}
simulations is also good, though the slope is slightly steeper in the
PESPH simulation. Also, compared to the \textit{nw} runs, the stellar
feedback in the \textit{ezw} runs only contributes a little extra hot gas
to massive groups, as also found in \citet{ford14} and \citet{liang16},
owing both to the small mass loading factors in massive galaxies and
the short wind recycling time-scales. The differences between the \textit{ezw} simulations seems to maximise around $M_h \sim 10^{13}M_\odot$, which is the key regime for galaxy transformation. However, these differences are small.

\subsection{Baryonic accretion}
The star formation history of the universe is closely linked to how galaxies
acquire their gas. The particle nature of SPH facilitates the straightforward study of
the accretion histories of galaxies. By tracking individual particles
through simulation outputs we are able to study the thermal histories of
accreted material that assemble into the simulated galaxies. We define an
accretion event to occur when a gas particle resides inside a resolved
SKID group ($M_{gal}>64m_{gas}$) when that SKID group did not contain
the gas particle in the previous output. For each simulation we have 75
outputs from $z=4$ to $z=0$. We subdivide the accretion into three major
channels \citep{keres05, keres09a, oppenheimer10, roberts18}. When a gas
particle is found in a resolved galaxy for the first time, we subdivide
the accretion as either hot pristine gas accretion (\textit{hot mode}),
if the maximum temperature $T_{max}$ of the particle reached at least $2.5\times 10^5$ K at any time before accretion, or cold pristine gas accretion (\textit{cold
mode}) if not. Once a particle accretes 
we reset its $T_{max}$ to zero and only allow
$T_{max}$ to change if the particle leaves the star-forming region. If
an accreted particle has been launched as a wind particle previously, we
define the accretion event as wind re-accretion (\textit{wind mode}). Note
that the accretion tracking program used here \citep{roberts18} also
classifies accretion into a few other sub-dominant channels (less than
10\%), but we omit them from the plot for the purposes of this paper and
do not consider them further.

\begin{figure*}
\centering 
\includegraphics[width=1.90\columnwidth]{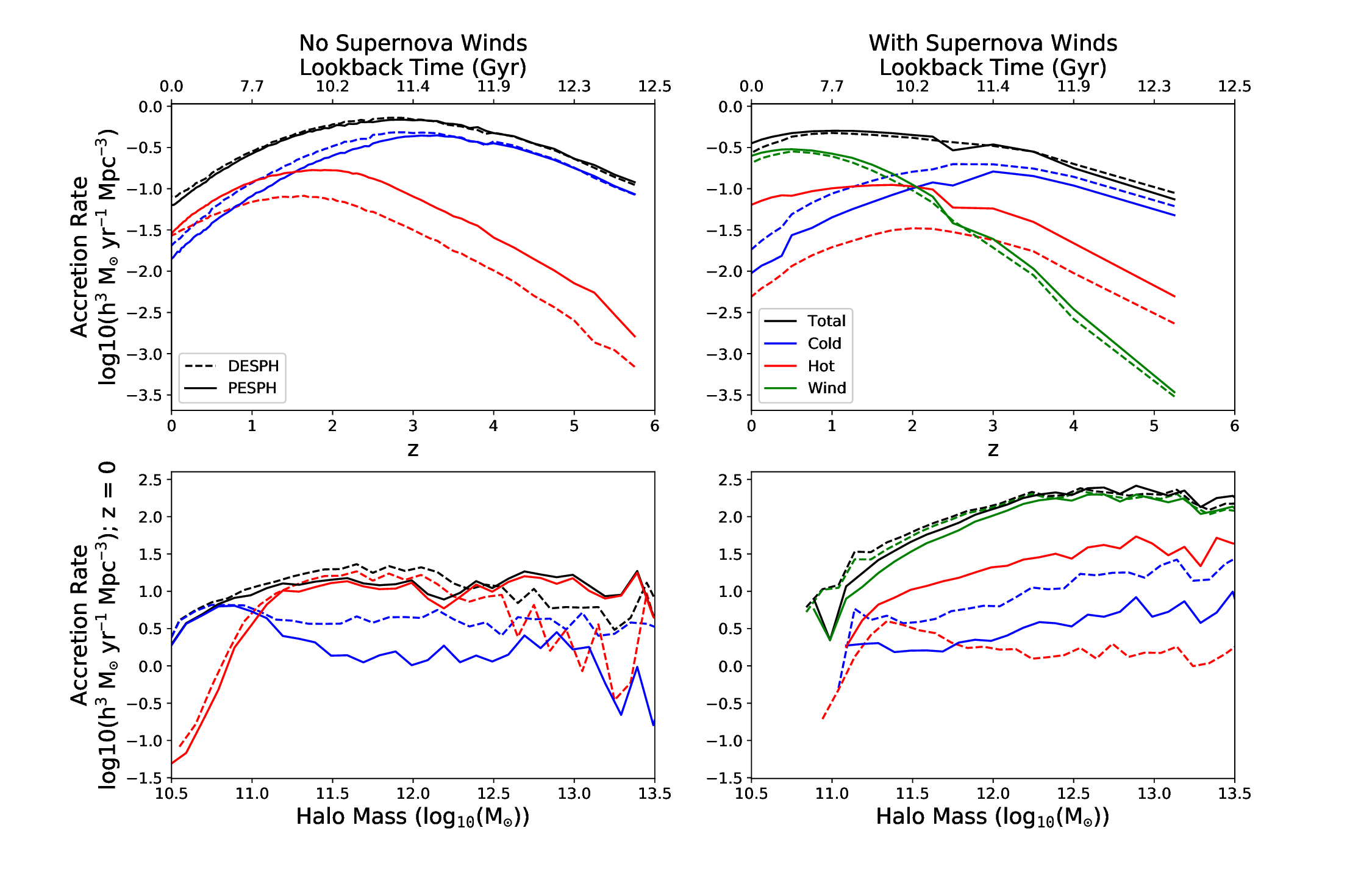}
\caption{\textit{Upper panels: }Global accretion rate as a function of
redshift from $z=6$ to $z=0$ in our simulations. We compare the two
\textit{nw} simulations in the left panels and the two \textit{ezw}
simulations in the right panels. The DESPH simulations and the PESPH
simulations are plotted as dotted lines and solid lines, respectively. We
divide the accretion into three channels defined in the text, namely, cold
mode pristine accretion (blue), hot mode pristine accretion (red) and (in \textit{ezw}
simulations where galactic winds are added using a sub-grid prescription)
wind re-accretion (green). The black lines in
each panel indicate the total accretion rates. \textit{Lower panels:
}Global accretion rate as a function of the host halo mass at $z=0$.}
\label{fig:acc_all} 
\end{figure*}

The top panels in Figure \ref{fig:acc_all} show how galaxies acquire their gas 
through the three major accretion channels: cold (blue), hot (red) and wind
re-accretion (green) across cosmic time for PESPH and DESPH with
(\textit{ezw}) or without (\textit{nw}) supernova winds. The most
significant change from our original DESPH formulation is that the new
PESPH formulation boosts the hot accretion rate by a large fraction at
nearly every redshift for both the \textit{nw} and \textit{ezw} models.
In the \textit{nw} simulations, galaxies accrete their gas through
either cold accretion or hot accretion. Since there are no winds, only
a small amount occurs owing to reaccretion of tidally stripped gas,
which we do not plot.
In the PESPH simulation, cold accretion is suppressed after
$z=3$ by a small fraction, but hot accretion is consistently enhanced
by half a dex. The total accretion rates, however, are very similar
between the two simulations. 

Figure \ref{fig:acc_all} (bottom panels) show how these process depend on
halo mass at $z=0$.  The accretion rate is
measured by counting the amount of accretion onto each halo during a
small time window.  Differences in the cold/hot
accretion rates are obvious only in haloes with $\log(M_h/M_\odot) > 11.0$, and the enhancement in hot mode accretion increases with halo mass.
In smaller mass haloes, galaxies acquire their baryons mostly from
cold streams that are never shocked above the temperature threshold
of $2.5\times 10^5$ K. The accretion rates in these haloes are hardly
affected by the SPH formulation. These haloes are generally devoid of a
hot corona, so that the hydrodynamical interactions between cold and hot
gas are unimportant. In the \textit{ezw} simulations, wind-reaccretion
starts to dominate after $z=2$. Although the hot accretion rate is strongly enhanced at $\log(M_h/M_\odot) > 10.5$, the impact of the new numerics on the dominant wind-reaccretion rate is insignificant, and the impact on cold accretion rates is small.

\begin{figure*} 
\centering
\includegraphics[width=1.90\columnwidth]{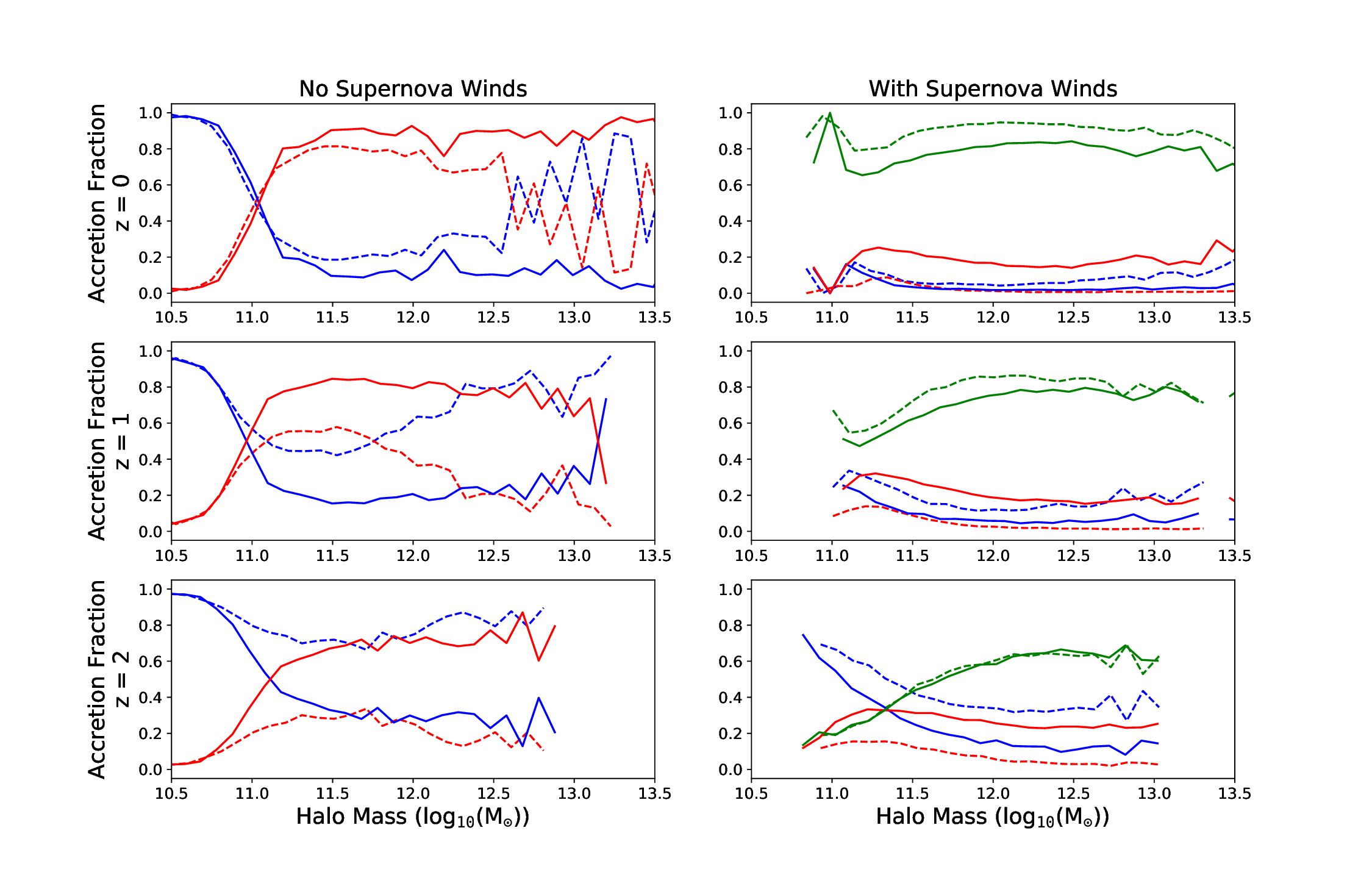}
\caption{In these plots we compare the fractional accretion rate
as a function of halo mass at $z=0, 1, 2$ (upper, middle and lower
panels). As in fig. \ref{fig:acc_all}, the left panels compare the two
\textit{nw} simulations and the right panels compare the \textit{ezw}
simulations. The dotted and solid lines in each panel correspond to
the DESPH and PESPH simulations, respectively.  } \label{fig:acc_z1}
\end{figure*}

Figure \ref{fig:acc_z1} shows how the instantaneous fractional
accretion rate through
the three channels depends on halo mass at different redshifts. 
There is a fractional enhancement of hot mode accretion in
both the \textit{nw} and \textit{ezw} simulations at all redshifts
when using the PESPH formulation. This numerical effect is most
prominent in the \textit{nw} simulations at $z=1$ and $z=2$: In the
original DESPH simulation, cold mode accretion dominates over hot mode
accretion in nearly all haloes; however in the PESPH simulation, hot
mode accretion starts to dominate in haloes with $\log(M_h/M_\odot) >
11$. In the \textit{ezw} simulations, wind-reaccretion still 
dominates at $z=0$ and $z=1$ and is not much affected by the SPH
algorithm. As in the \textit{nw} simulations, PESPH significantly
enhances hot mode accretion, making it comparable to cold mode
accretion in intermediate to massive haloes. The results for DESPH are similar to those of \citet[fig. 8]{keres09a} as expected, while those for PESPH are closer to the earlier results of \citet[fig. 6]{keres05}, though with a cold-to-hot transition shifted downward by $\sim 0.3$ dex in halo mass.

In previous cosmological simulations that use the traditional DESPH
formulation \citep{keres09a}, the cold dense gas clumps that form either
by spuriously numerically enhanced thermal instabilities
or stripping contribute a considerable amount of pseudo cold accretion as
they fall onto the central galaxy. The PESPH formulation allows efficient
mixing between these cold structures and their hot surroundings, prevents these cold clumps from falling onto the galaxies, thereby
suppressing cold accretion. Besides, the enhanced mixing lowers
the thermal energy of the hot halo gas, which can, therefore, cool more
efficiently, allowing for more hot mode accretion. This provides a possible
explanation for the differences between the accretion rates in the DESPH
and PESPH simulations. The new artificial viscosity reduces artificial
heating from numerical noise in the velocity field where the flow is non-convergent, but increases heating
by capturing shocks more accurately. However, it is challenging to measure
its net effects quantitatively. The increase in hot gas fractions for PESPH shown in Figure \ref{fig:fhot} probably indicates that some of the increase in hot mode accretion, which comes at the expense of a decrease in cold mode accretion (Figure \ref{fig:acc_all}), owes to more shock heated gas.

The PESPH simulations retain the key qualitative findings of our earlier papers \citet{keres05, keres09a, oppenheimer10}: in \textit{nw} simulations, cold mode accretion dominates galaxy growth at low halo masses and high redshifts; in \text{ezw} simulations, wind recycling dominates galaxy growth in massive haloes at low redshifts. However, hot mode accretion rates are sensitive to the difference between DESPH and PESPH, and these differences can shift the redshifts or masses at which hot accretion comes to dominate over cold accretion. This result is consistent with findings from AREPO simulations \citet{nelson13}.

\begin{figure} 
\centering
\includegraphics[width=1.00\columnwidth]{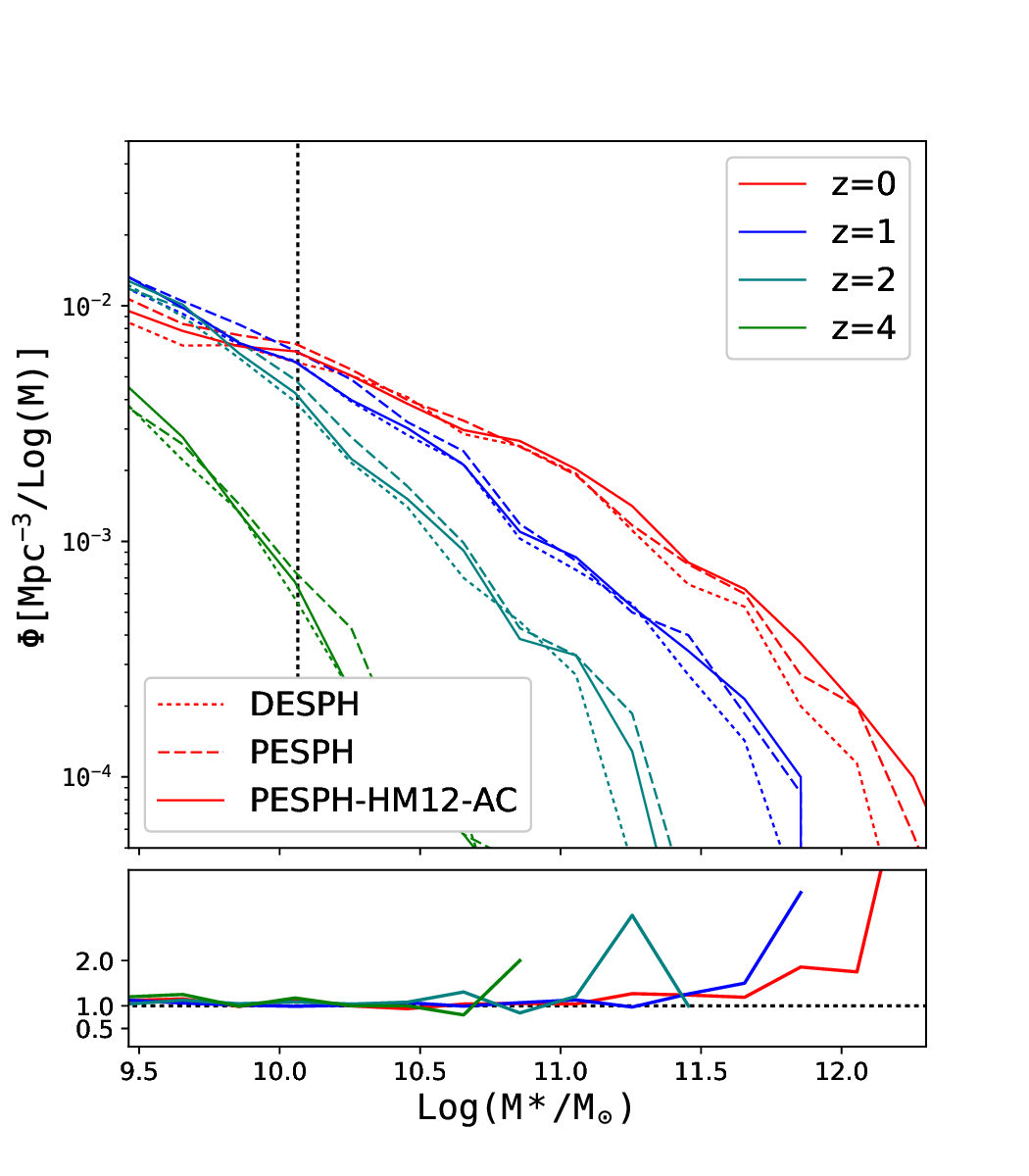} \caption{The
upper panel shows the galactic stellar mass functions (GSMFs) at
$z=0$(red), $z=1$(blue), $z=2$(teal), and $z=4$(green). The dotted,
dashed, and solid lines correspond to the DESPH, PESPH, and PESPH-HM12-AC
simulations. The dotted vertical line shows half of the resolution limit
(32 gas particles) for SKID galaxies. The bottom panel shows the ratio
of the number densities of galaxies from the PESPH-HM12-AC simulation to the
DESPH one at each stellar mass bin. The same colour scheme is used as in
the upper panel.  } \label{fig:gsmf} 
\end{figure}

\subsection{Galaxies} In Figure \ref{fig:gsmf} we show the galactic
stellar mass functions (GSMFs) at four redshifts. Despite the differences
in the numerical schemes and the cooling model, the GSMFs in low mass
galaxies with $M_* < 10^{10.5}M_\odot$ are nearly identical for all three
simulations compared here at any redshift. In the most massive galaxies,
the numerical scheme does make a small yet noticeable difference. Both
PESPH and PESPH-HM12-AC use the pressure-entropy formulation and produce
slightly more massive galaxies than DESPH, but the differences between these two
simulations are quite small. Even though the growth of massive galaxies is
most likely dominated by physical assumptions rather than the numerics,
this change at the high mass end could provide further constraints on
a proper quenching model to suppress star formation in these massive
haloes. As
Figure \ref{fig:smhm} shows, the star formation in low mass galaxies
is strongly regulated by our kinetic feedback scheme. The results of
\citet{oppenheimer10} show that a slight change in the feedback model
leads to very different stellar mass functions. Here our results confirm
that feedback is crucial to explaining the observed stellar mass function
regardless of the numerical errors introduced by the traditional SPH
formulations.

The results on the GSMFs are consistent with our previous results
on the accretion rates and the stellar mass - halo mass relation. The stellar mass growth depends on the total accretion onto the galaxies. Even though the hot mode accretion is considerably enhanced by the new SPH techniques (Figure \ref{fig:acc_all}), it is compensated by the reduced cold accretion. The total accretion rate, which is dominated by cold mode accretion at high redshifts and by wind re-accretion at low redshifts, hardly changes.

By comparing the traditional SPH and ANARCHY SPH, \citet{schaller15} also find that the GSMFs are very insensitive to PESPH and other numerical improvements. However, they find that ANARCHY SPH produces slightly less massive galaxies at z = 0, opposite to our findings. Though the ANARCHY scheme is very similar to our fiducial SPH scheme, they use a feedback prescription that is very different from us. It is not straightforward to determine how different feedback schemes are specifically affected by even the same numerical changes.

Besides the stellar mass functions, we have
examined the global star formation histories in our simulations, as well
as the properties of individual galaxies, such as the relation between
the specific star formation rate (sSFR) and stellar mass, the gas phase
mass-metallicities relation, and the gas fraction as a function of 
galactic mass. None of these results are strongly affected by either
the numerics or the non-equilibrium cooling model/HM12 background. 
Detailed comparisons can be found in Appendix C and Appendix D.

\section{Intergalactic and Circumgalactic Medium}
\begin{figure*} 
\centering
\includegraphics[width=1.9\columnwidth]{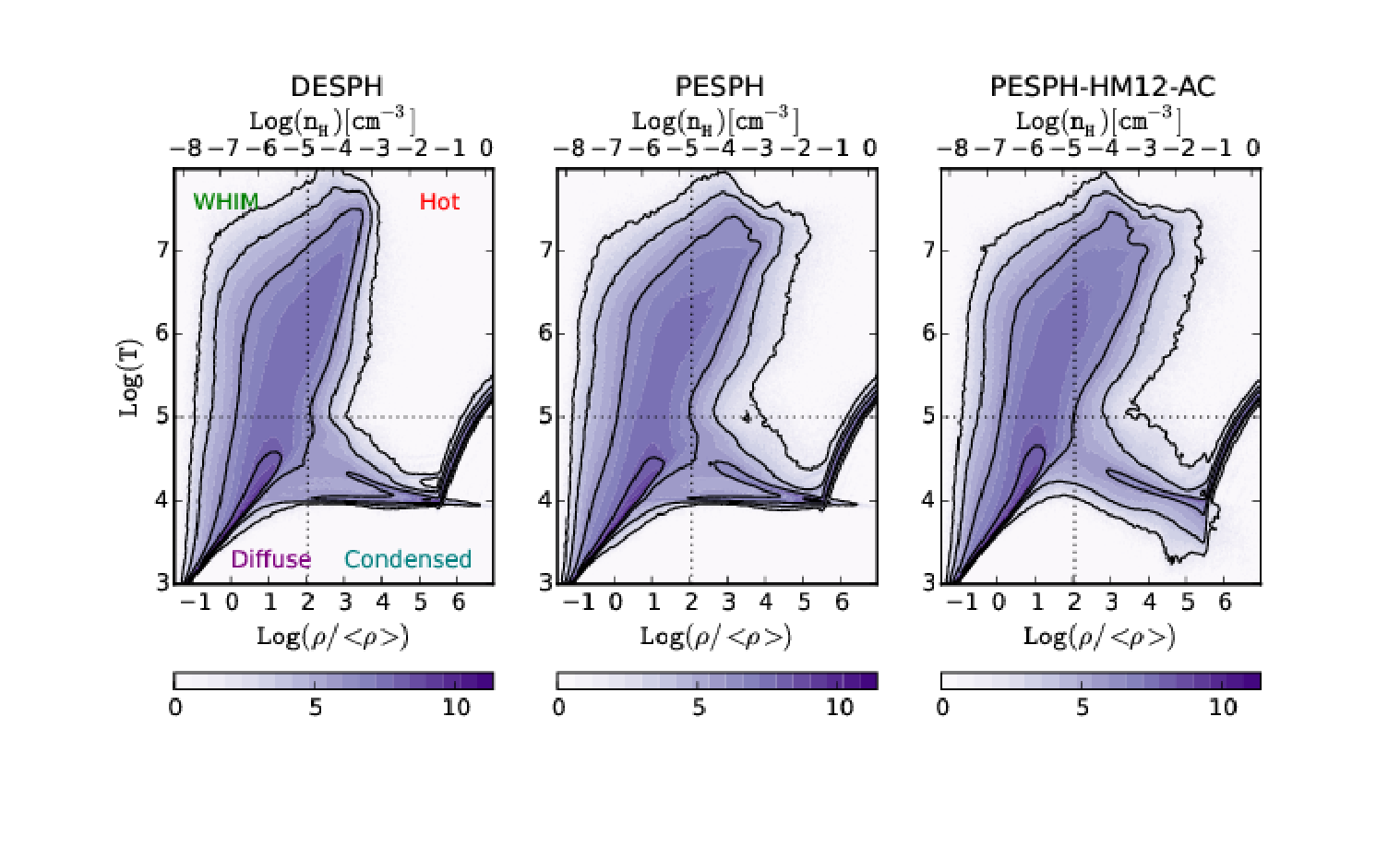}
\caption{Density-temperature phase diagrams from the DESPH, PESPH and
PESPH-HM12-AC simulations at $z=0.25$. The grey scale indicates the gas
particle number density. The dotted lines overplotted in each panel divide
the baryons into different phases following \citet{dave10}. The upturn
starting from $n_H > 0.13\ \mathrm{cm^{-3}}$ results from our assumption
of an effective equation of state for multi-phase interstellar medium
gas \citep{springel03}. Despite noticeable differences in a few regions,
the distribution of gas particles in the three simulations are quite
similar, indicating that the global gas properties are insensitive both to
changes in the SPH algorithm and the cooling physics.} \label{fig:rhot} 
\end{figure*}
In this section we restrict our discussion to the simulations with galactic winds (ezw). Relative to our older DESPH prescriptions, we examine the impact of the pressure-entropy formulation, the \citet{wiersma09} metal cooling and the HM12 UV background (PESPH-HM12), and the changes to the artificial viscosity and the artificial conduction (PESPH-HM12-AC).

The thermodynamic properties of baryons in the simulated volume
can be conveniently studied by looking at the phase diagram (Figure
\ref{fig:rhot}), in which the SPH particles are binned in density
- temperature space. Following \citet{dave10}, we divide all gas
particles into five categories depending on their locations in the
phase diagram. To separate regions in the diagram, we choose a constant
temperature threshold of $T_{th} = 10^5$ K and a redshift dependent
overdensity threshold of $\delta_{\rho}$ defined by \citep{kitayama96}:
\begin{equation} \delta_{\rho}=6\pi^2(1-0.4093(1/f_\Omega-1)^{0.9052})-1
\end{equation} where \begin{equation} f_\Omega =
\frac{\Omega_m(1+z)^3}{\Omega_m(1+z)^3+(1-\Omega_m-\Omega_\Lambda)(1+z)^2+\Omega_\Lambda}
\end{equation} The threshold reflects the overdensity at the boundary
of a virialised halo and roughly separates gas within dark matter haloes from
that outside them. The thresholds are shown as dotted lines in Figure
\ref{fig:rhot}. Though the classification is a simple one, tests have
shown that the gas particles that fall in each region represent different
baryonic environments \citep{dave10}. The lower left of the diagram
consists mainly of diffuse primordial gas. Most of these gas particles
lie on a well defined curve in the phase diagram, which is established
by a balance between adiabatic cooling and photoionisation heating. A
fraction of gas particles are shock heated when they collapse into the
gravitational potential of dark matter sheets and filaments and are driven 
into warm-and-hot ionised gas outside
of haloes (upper left) or fall into dark matter haloes and become 
hot halo gas (upper right). Radiative cooling
later plays a critical role in the further condensation of gas into the
condensed region (lower right) where SF can occur. In addition, some gas
goes straight from the diffuse to the condensed region, i.e. cold mode 
accretion \citep{keres05, keres09a}. The upturn in the
densest region owes to multi-phase ISM particles with densities above
the star formation density threshold of $n_H = 0.13\ \mathrm{cm^{-3}}$,
which follow an effective equation of state. All ISM particles are
counted as condensed gas even if their temperatures pass $T_{th}$. The
fifth baryonic phase is stars.

The global distributions are quite similar in all three simulations,
indicating that the phase structure of baryons on a global scale is not
significantly affected by the hydrodynamical formalisms or the radiative
cooling model. In the hot dense region that represents shocked halo gas,
simulations with the PESPH formulation extends the hot gas to higher
densities, lowering their entropy. Since this hot gas is most responsible
for X-ray emissivity, the enhanced density could lead to a considerable
boost in X-ray luminosities in PESPH haloes at given mass (see section 6.4). The region
near the point where gas phases intersect is also slightly affected. The
phase diagrams of the PESPH and the fiducial PESPH-HM12-AC simulations
resemble each other except in the condensed gas region. In the left and
middle panel, the locus of particles at $T \sim 10^4$ K consists mostly
of enriched gas particles for which the CIE metal line cooling becomes
negligible at that temperature. In the PESPH-HM12-AC simulation, on
the other hand, photoionised electrons from heavy elements allow these
particles to cool further, until it establishes a density dependent
thermal equilibrium. However, photoionisation suppresses metal line
cooling in the warm, less dense gas, leading to noticeable differences
at $n_H = 10^{-5}$ to $10^{-4}\ \mathrm{cm^{-3}}$ and $T = 10^{4}$ to 
$10^{5}\ K$.

\begin{figure} 
\centering
\includegraphics[width=1.00\columnwidth]{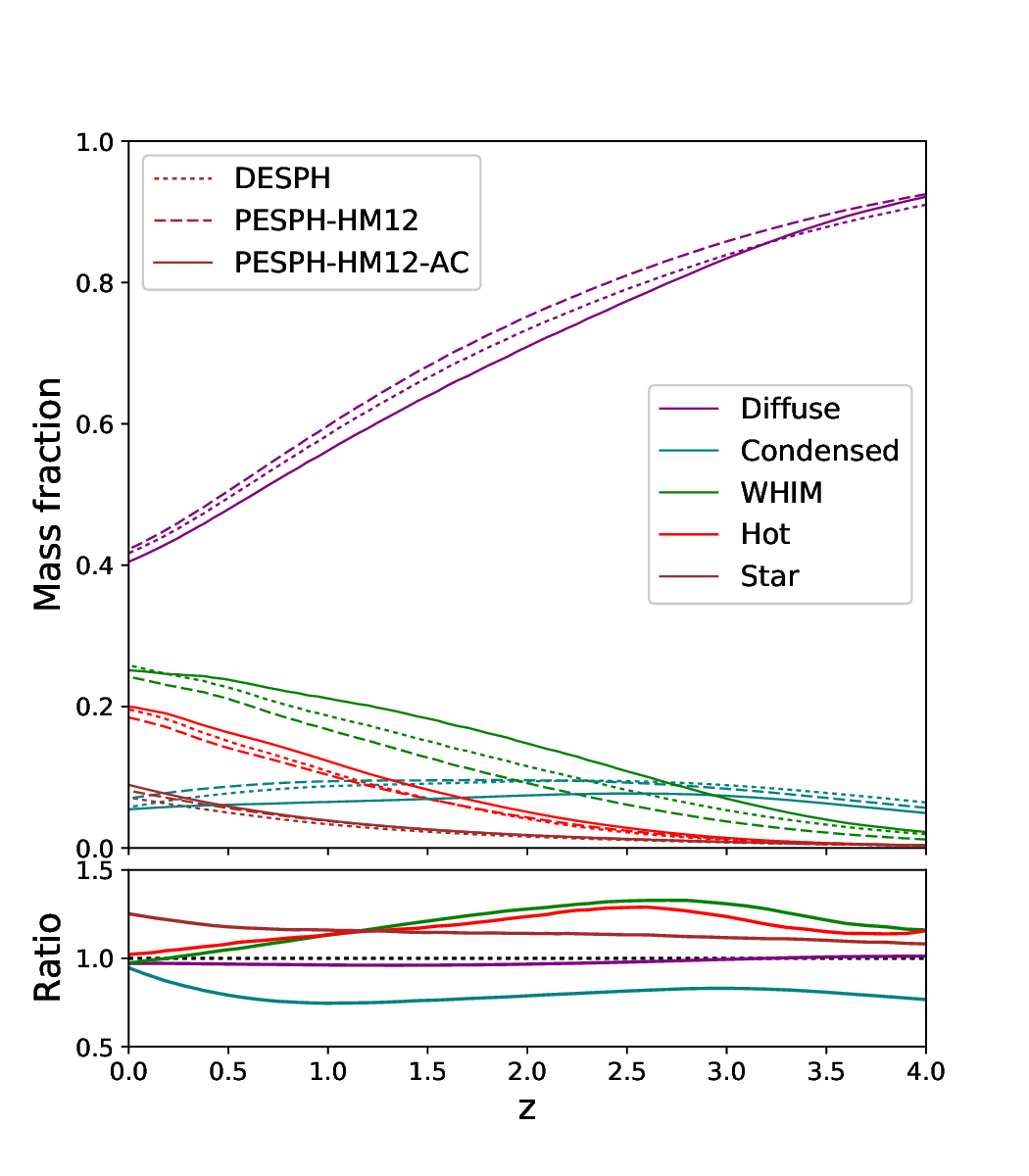} \caption{The
upper panel shows the evolution of the different baryonic phases. We plot
the mass fractions in each phase as a function of redshift for DESPH,
PESPH and PESPH-HM12-AC simulations as dotted, dashed and solid lines,
respectively. The purple, green, red, teal and brown lines represent
cold diffuse IGM gas, warm and hot ionised medium, hot ionised halo gas,
dense cold gas in galaxies and gas locked into stars, respectively. The
definition of each phase is given in \S6. The amount of gas in each
phase in the different simulations is very similar at all times. The
bottom panel shows the ratio of the mass fraction of each phase in the
PESPH-HM12-AC simulation to the DESPH one, as a function of redshift. The
same colour scheme is used as in the upper panel.  } \label{fig:fphase}
\end{figure}

For a more quantitative comparison, Figure \ref{fig:fphase} shows
how the fraction of baryons in each phase changes over cosmic time. The
mass fractions of the four gas phases plus the stellar component from
the three simulations are shown as a function of redshift. Baryons start
in the diffuse phase and enter the other phases as structure grows
\citep{dave10}. A striking agreement is seen between these different
simulations. The total amount of hot gas in PESPH is only slightly
larger than in the DESPH, despite the fact that there is a noticeable
increase in the hot gas at higher densities in the PESPH simulation
(Figure \ref{fig:rhot}). The fiducial PESPH-HM12-AC simulation is the
most different from the other simulations. The diffuse gas fraction is
lower since $z \sim 2.5$ when the WHIM and hot fraction has increased
by nearly 30\% and declines thereafter. The WHIM and hot phase in our
simulations are formed primarily from shock heated gas, but also has
contributions from galactic winds. These changes are most likely caused
by the \citet{cd10} viscosity, which reacts faster to converging flows
than the M\&M97 viscosity. It makes the new viscosity more sensitive
to accretion shocks around the filaments and the wind particles as soon
as they recouple hydrodynamically. Indeed, when we track the dynamical
histories of individual particles in some smaller test simulations we find
that the wind particles slow down more quickly with the new viscosity.

Observationally the intergalactic medium is often
probed using quasar absorption line spectra. To mimic this technique, we
take sightlines through our simulation box and generate absorption spectra
using SPECEXBIN. A thorough description of this method can be found in
\citet{oppenheimer06}, we only give a brief summary here. We take lines of
sight through the periodic box, and divide them into tiny redshift bins,
so that each ``pixel'' of the spectra corresponds to one redshift bin. In
each bin, the optical depths of various species are computed based on
the smoothed local physical properties such as density, temperature,
metallicity and background radiation. These properties are evaluated by
smoothing over all nearby SPH particles within the smoothing kernel. %
We adjust the HI optical depth later by matching the evolution of Lyman
$\alpha$ flux decrements to observations following \citet{dave10}. We
multiply all the optical depths by a redshift independent correction factor of 0.62 to DESPH and PESPH results and
a factor of 0.31 to PESPH-HM12 and PESPH-HM12-AC results. This is equivalent to enhancing the background ionising flux. The value chosen for the simulations using the HM01 background is close to that used in \citet{dave10}, which also assumes the HM01 background. However, when we use the HM12 background, a larger correction is needed to match the flux decrement constraints, indicating that ionising photons from the HM12 background are insufficient to ionise the Lyman $\alpha$ forest at low redshift. This is consistent with the findings of \citet{kollmeier14}.

To understand the absorption on different
scales, we generate two kinds of absorption spectra. First, we take 70
random lines of sight, each of which wraps around the simulation box a few times until it covers a redshift range from $z=0$
to $z=2$. These spectra are used to normalise the radiation background,
study the evolution of elements, and study the global column density
distributions of several species. We normalise the background UV
field intensity to match the observed mean Ly$\alpha$ opacity. Second, we take short targeted lines of sight surrounding
haloes, following \citet{ford13}. We select 250 haloes that fall in the
mass range $11.75 < \log(M_h/M_\odot) < 12.25$ out of each simulation at
$z=0.25$. For each halo, we take short sightlines that penetrate the halo
at different radii. We generate four spectra for each of the 12 impact parameters
ranging from 10 kpc to 300 kpc. This procedure results in 12,000 short
spectra for each simulation. We apply COS instrumental S/N and resolution
to all spectra, long or short, so that our statistics are comparable
to those obtained from real COS spectra and also directly comparable
to \citet{ford13}.

We fit Voigt profiles to each spectrum using AUTOVP \citep{dave97}. Each
ion is fit separately. AUTOVP then outputs the column density and
equivalent width (EW) for each fitted feature. We combine individual
metal lines into systems if their line centres are closer than 100 km/s
in velocity space, by adding the column densities and EWs to the largest
feature. For short sightlines surrounding the haloes, we focus our study on
features within $\Delta v \pm 300 km/s$ of the central galaxy. This range
covers the vast majority of absorption within the halo \citep{ford13}.

\subsection{The Lyman-alpha forest} 

The Ly$\alpha$ forest is a distinct feature in high redshift quasar
spectra that arises from absorbers that lie along the line of sight to
distant quasars. The Ly$\alpha$ forest has been an important observational
constraint on the $\Lambda$CDM cosmology and provides valuable insights on
structures in the intergalactic and circumgalactic medium \citep{kollmeier03, dave10, peeples10a, peeples10b, kollmeier14}. \citet{dave10}
show that the statistics of Ly$\alpha$ absorbers is insensitive to the
feedback prescription, despite the strong dependence of the global star
formation history and the halo enrichment on feedback. In this section
we study the robustness of the predicted Ly$\alpha$ column density
distribution to different numerical and physical schemes.

\begin{figure} 
\centering
\includegraphics[width=1.00\columnwidth]{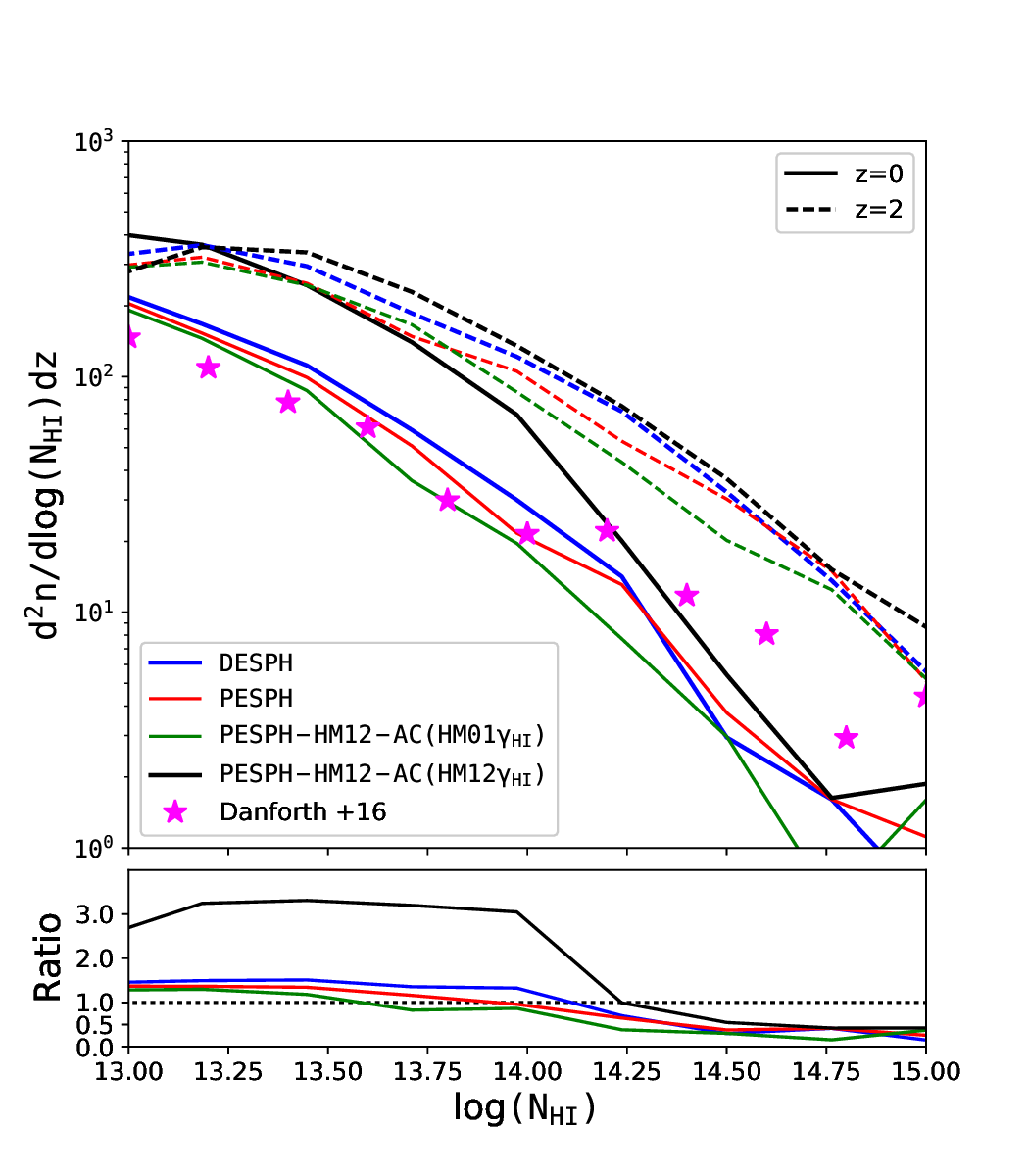} \caption{The
\textit{upper} panel compares the column density distributions (CDDs)
for Ly$\alpha$ at $z=0$ and $z=2$. Blue and red lines present the
CDDs from the DESPH and PESPH simulation with the HM01 background. The
black lines show the CDD from the fiducial PESPH-HM12-AC with the HM12
background. Lastly, we recalculate the CDD for this run using the HM01
background for comparison (green lines). The $z=0$, and $z=2$ CDDs are
plotted as dashed and solid lines, respectively. The turnover at $N_{HI}
\sim 10^{13}\ \mathrm{cm^{-2}}$ shows the incompleteness in recovering
under resolved lines assuming COS instrumental S/N and resolution. Data
from \citet{danforth16} are plotted as magenta stars. The \textit{lower}
panel shows the ratio between the measured CDD at $z=0$ from simulations
and the \citet{danforth16} data.  } \label{fig:cdd_hi} 
\end{figure}

Figure \ref{fig:cdd_hi} compares the HI column density distributions
derived from Ly$\alpha$ absorbers at $z=0$ and $z=2$. We plot
$d^2n/(d\log(N)dz)$ for better readability. The column densities are obtained
from fitting Voigt profiles to all Ly$\alpha$ lines identified by AUTOVP over 70 random lines
of sight that have an equivalent width broader than
$0.03 \AA$ within the redshift ranges $z=0-0.2$ and $z=1.8-2.0$, covering
a total $\Delta z = 14$. Qualitatively all four simulations show similar CDDs at both redshifts. 

All three simulations (DESPH, PESPH, PESPH-HM01-AC) that use the same
UV background (HM01) produce similar HI column density distributions
at both redshifts, suggesting that the Ly$\alpha$ statistics are robust to different SPH schemes, despite the fact that hydrodynamical
instabilities are much better captured in the new code. This is consistent
with previous results that the baryonic structures are largely unaffected
by the numerics. In detail, the PESPH-HM12-AC using the HM01 background yields a slightly lower CDDs at both redshifts. However, comparing the black line and the green line for
$z=0$ indicates that changing the background from HM01 to HM12 increases
the number density of low column HI absorbers by roughly 0.5 dex.

Our result supports the claim that the photoionisation rate derived
in HM12 is insufficient to explain the abundance of low redshift HI
absorbers \citep{kollmeier14}. We compare our derived CDDs to the new
observational constraints from the HST/COS survey \citep{danforth16} and
find that the new fiducial simulation PESPH-HM12-AC still overproduces HI absorbers by a factor of 3 at low column densities. In order to match the observed distribution at $z=0$, we need to artificially boost the UV flux by a factor of $\sim 5$, a value that is consistent with \citet{kollmeier14}. However, if we apply the HM01 background
in our post-processing measurement, the CDD is very close to those from
the DESPH and PESPH simulations. Therefore, the updated UV background is
solely responsible for the enhancement of HI absorption in the our new
fiducial simulation. Neither the SPH formulation nor the new cooling
model has a strong impact on the Ly$\alpha$ statistics.

\subsection{Metals} 

\begin{figure*} 
\centering
\includegraphics[width=1.9\columnwidth]{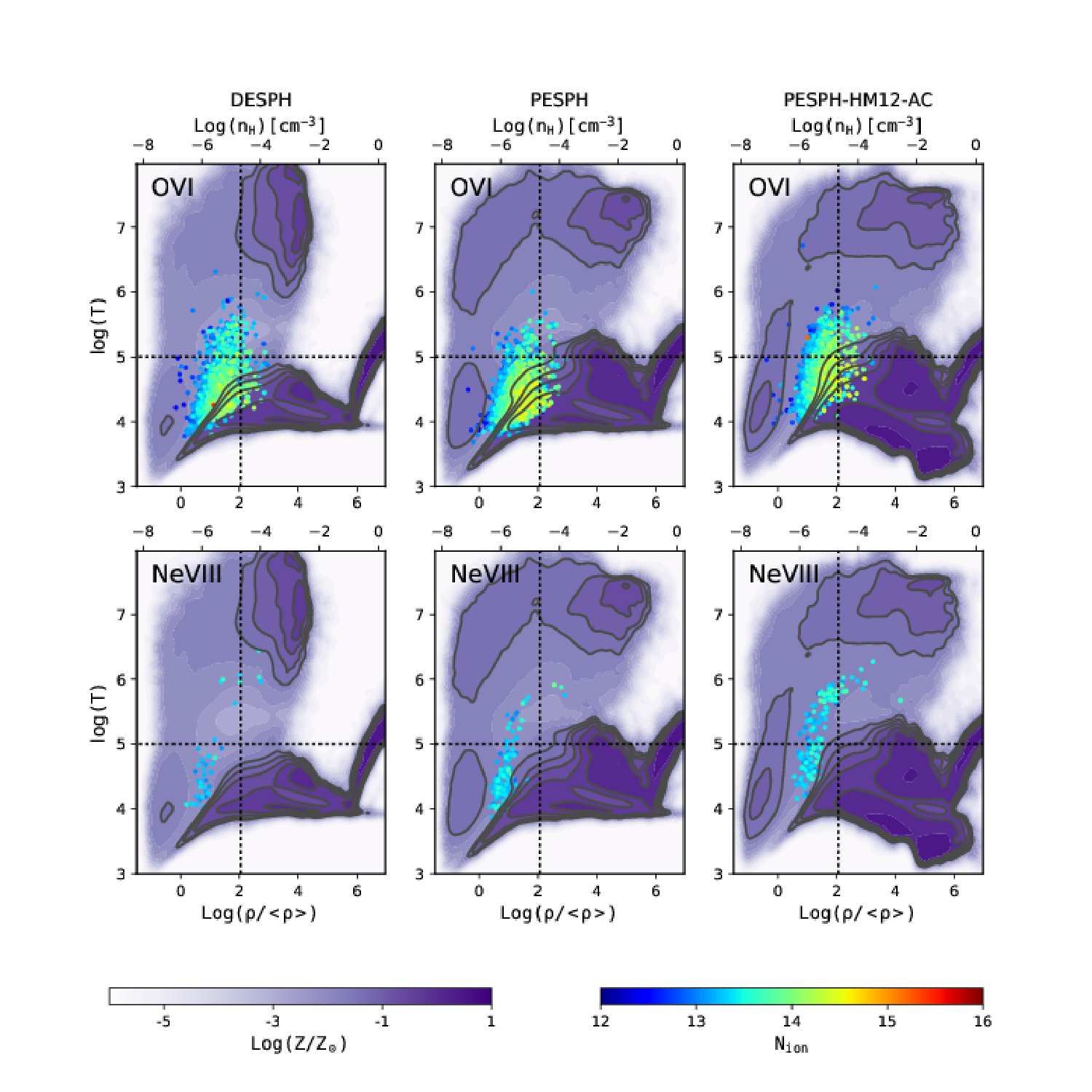}
\caption{From left to right, the metal distribution in the
density-temperature plane from the DESPH, PESPH and PESPH-HM12-AC
simulations. The purple colour scale indicates the mass weighted
average metallicity compared to solar. The solid lines divide the
gas into four phases as defined for Figures \ref{fig:rhot} and 
\ref{fig:fphase}. Over plotted dots represent OVI (\textit{upper panels}) and NeVIII (\textit{lower panels}) absorbers identified
from $z=0$ to $z=0.5$ from 70 sightlines. The colour scale, which
is identical in all panels, indicates the column densities of these
absorbers, spanning from $10^{12}\ \mathrm{cm^{-2}}$ to $10^{16}\
\mathrm{cm^{-2}}$. The strong absorbers (redder in colour) have been
stacked on top of weaker absorbers. Several contours lines are stressed
for better visualisation.  } \label{fig:metal_rhot_ovi_neviii} 
\end{figure*}

Figure \ref{fig:metal_rhot_ovi_neviii} compares the distribution of enriched
gas in density-temperature phase space. The density and temperature
dependence of metallicity is closely related to the outflows and
encodes information of how outflows heat and enrich the gas in various
phases. Though the metals locked within galaxies barely differ between
simulations (see Figure \ref{fig:mzr} in Appendix E),
the metal distributions in the gas
phases are distinct. Specifically, the WHIM gas is more metal enriched
in both the PESPH and PESPH-HM12-AC simulations, and the metals in the
cool, condensed gas are more extended towards the hotter, less dense
region. Satellite galaxies that enter hot haloes are more vulnerable to
disruption owing to the enhanced hydrodynamical instabilities and mixing
efficiency between fluid interfaces. In addition, shocks around filaments
are better resolved, making more WHIM, metal rich gas. Furthermore,
galactic winds are enhanced in massive galaxies in the PESPH simulation,
further facilitating the mixture of enriched gas and metal-poor gas.

To compare the physical conditions from which observed absorption arises,
we show the locations of OVI and NeVIII absorbers that we have studied previously
\citep[e.g.,][]{oppenheimer12} in the phase diagram. Most OVI absorbers
still reside in diffuse gas where photoionisation dominates. The improved
hydrodynamics leads to an increased number of strong absorbers in the
condensed gas, in which the metal content now has a broader distribution. Similarly, there are more NeVIII absorbers in the PESPH and the PESPH-HM12-AC simulations, owing to higher metallicity in the warm gas.

\begin{figure*} 
\centering
\includegraphics[width=1.75\columnwidth]{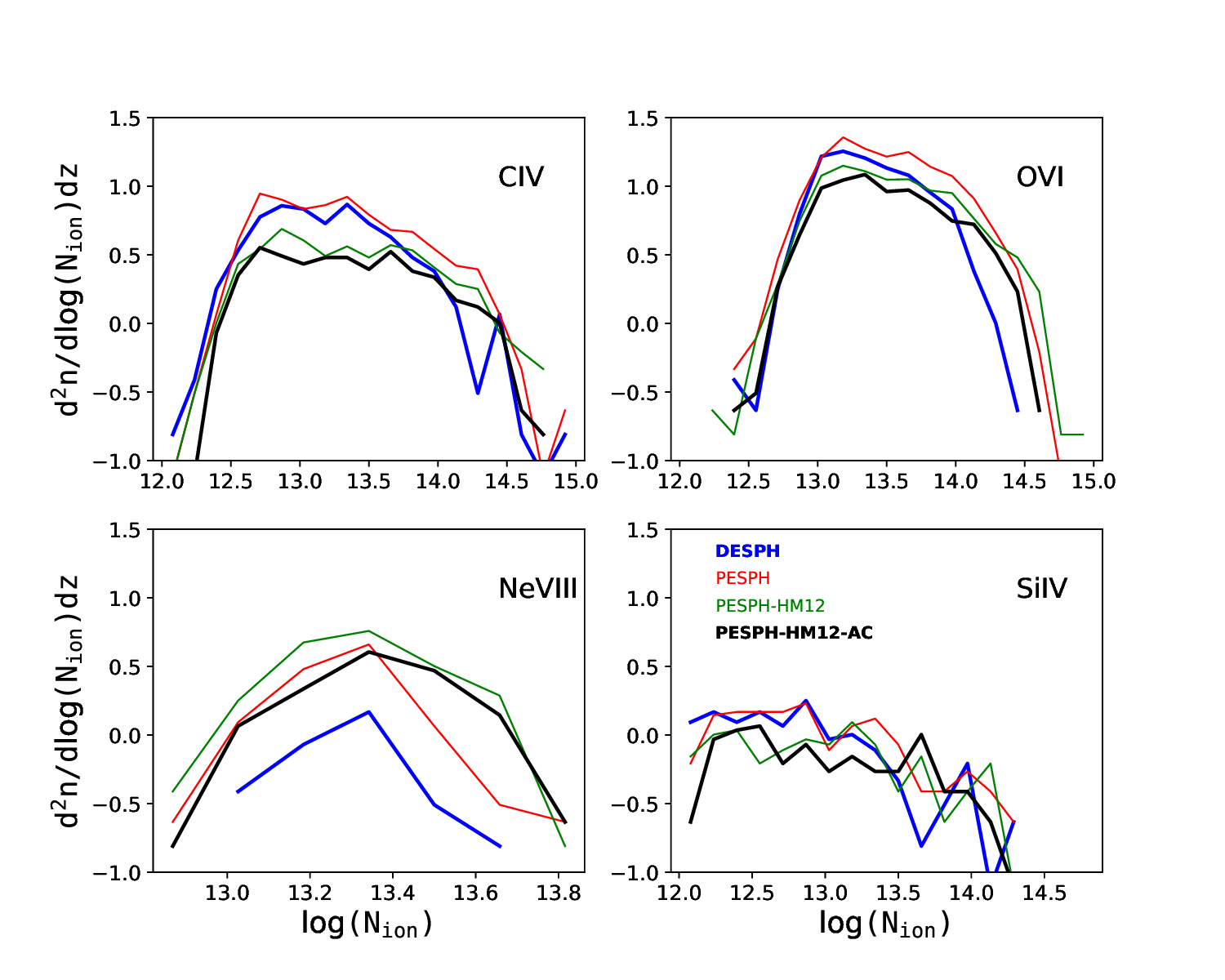}
\caption{Comparison of the column density distributions of CIV, OVI,
SiIV and NeVIII as labelled. Blue, red, green and black lines represent
the CDDs from the DESPH, PESPH, PESPH-HM12 and PESPH-HM12-AC simulations,
respectively. Nearby components fitted by AUTOVP that are within 100 km/s
are combined into systems whose column densities and equivalent widths
are summed over each component.  } \label{fig:cdd_metals} 
\end{figure*}

The column densities for metal ions are obtained in a similar
way as for HI, except that the spectral lines are sampled from a
wider redshift range, from $z=0$ to $z=0.5$ for each ion. Figure
\ref{fig:cdd_metals} compares the CDDs of several metal ions and
shows the differences between the simulations. Changing SPH formulation alone
(PESPH) leads to more absorption for both OVI and NeVIII. Changing the cooling prescriptions and the
background boosts NeVIII absorption
even further. Adding the new Cullen-Dehnen viscosity somewhat increases the high
column OVI and slightly reduces the low column NeVIII absorption. The most important change is the factor of $5 \sim 10$ boost in the CDD of NeVIII absorbers between DESPH and our fiducial calculation, which implies much better prospects for detecting this high-ionisation line.  As shown in Figure \ref{fig:metal_rhot_ovi_neviii}, the increase owes to more metals in hot haloes.

\subsection{Absorption in haloes} The absorption
line systems in the spectra of bright distant objects probe the internal
structures within a gaseous halo and, therefore, provide important
constraints on the galactic winds that change the chemical
and thermodynamic structures of simulated haloes. In this section,
we create mock observations of halo gas surrounding the simulated
galaxies at $z=0.25$ and examine whether predictions such as those
in \citet{ford13, ford16} are sensitive to the numerics.

\begin{figure} 
\centering
\includegraphics[width=1.00\columnwidth]{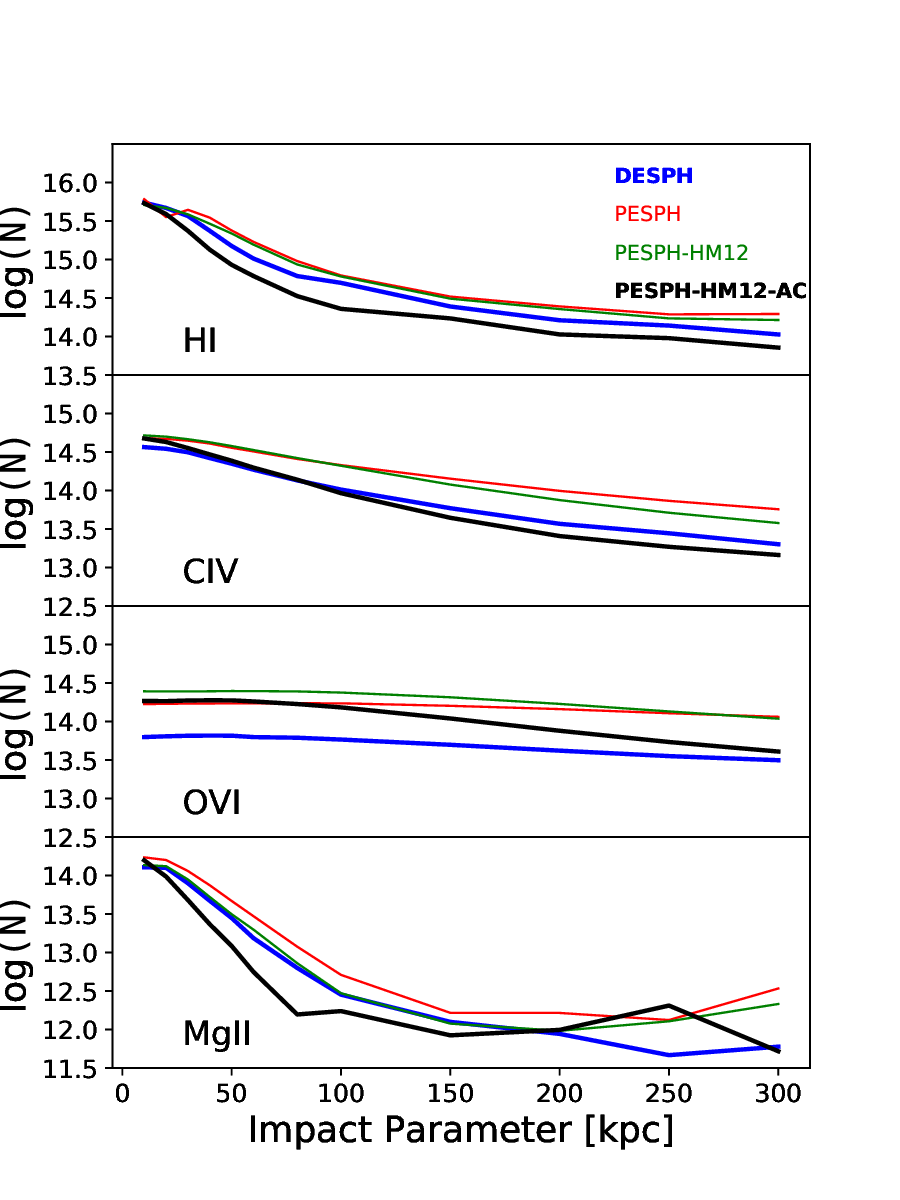}
\caption{The averaged absorption profiles of 250 selected haloes from each
simulation at $z=0.25$, with $11.75 < \log(M_h/M_\odot) < 12.25$. The
absorption is traced by the total column densities of several species
as a function of impact parameter ranging from 10 kpc to 300 kpc. Blue,
red, green and black lines show results from the DESPH, PESPH, PESPH-HM12
and PESPH-HM12-AC simulations, respectively.  } \label{fig:avgcd_halo}
\end{figure}

Figure \ref{fig:avgcd_halo} shows the extension of absorption within
$\log(M_h/M_\odot)\sim 12$ haloes at $z=0.25$. At each impact parameter,
we average over the column densities of all identified lines that
are broader than $EW = 0.03\ \mathrm{\AA}$ using over $\sim 1000$
lines of sight in 250 selected haloes. The $\log(M_h/M_\odot) \sim 12$
haloes constitute a representative sample that is well resolved for
studies of absorption. We choose upper limits of $10^{16}\ \mathrm{cm^{-2}}$ for HI and $10^{15}\
\mathrm{cm^{-2}}$ for other species. Column densities above these limits
are reset to the limit value before being added. For MgII, the absorption
drops dramatically within 100 kpc, while for ions with high ionisation
potentials, such as CIV and OVI, the strength of absorption only mildly
decreases as the impact parameter increases. The rise in MgII absorption
at large radii is probably contamination by neighbouring galaxies.

Changing the
SPH formulation to PESPH results in stronger absorption in all the species
plotted here, with only a slight dependence on the impact parameter. The increase in the high ions such as CIV and OVI owes to more hot metals in the halo gas as shown in Figure \ref{fig:metal_rhot_ovi_neviii}. The
absorption of HI within these massive haloes is more sensitive to the
choice of SPH algorithm than we previously saw for the global HI column
density distribution (Figure \ref{fig:cdd_hi}), indicating that the impact
of the improved schemes is stronger within these massive haloes. Changing
the cooling and UV background (PESPH-HM12) has little further effect
except for MgII, which becomes smaller at all radii. However, switching
to the new viscosity lowers the CIV at all impact parameters almost to the
original DESPH values, and also reduces the OVI columns at large radii.

In summary, neither the new PESPH formulation nor the metal cooling model significantly changes our predictions for the global Lyman $\alpha$ statistics at z = 0 and z = 2. Using the HM12 background in post-processing calculations produces many more Lyman $\alpha$ absorbers in the column density range of $10^{13}\ \mathrm{cm^{-2}} < N_{HI} < 10^{14}\ \mathrm{cm^{-2}}$, consistent with the findings from \citet{kollmeier14}. The metal absorption line statistics are more sensitive to our algorithms. The PESPH formulation alone results in more CIV absorptions at $N_{CIV} > 10^{13.5}\ \mathrm{cm^{-2}}$ and significantly more OVI and NeVIII absorbers at nearly all column densities. Adding the new metal cooling model does not further change the results, except for increasing the number of NeVIII detections even more.

\subsection{X-ray galaxy groups}
Both Fig.\ref{fig:rhot} and Fig.\ref{fig:metal_rhot_ovi_neviii} have shown qualitative differences in the dense hot gas, which is closely related to X-ray emission. To study whether our predictions on the X-ray properties of galaxy groups \citep{dave08,liang16} still hold, we derive X-ray luminosity weighted quantities following their method. The groups for this study are selected from the SO haloes that contain more than 8 SKID galaxies above the $6.0\times10^{9}M_\odot$ baryonic mass resolution limit. First, we generate the X-ray spectrum for every SPH particle based on its density, temperature, metallicity and the local radiation background using the Astrophysical Plasma Emission Code (APEC) models \citep{smith01}. From each output X-ray spectra, we integrate the flux from 0.5 keV to 10 keV to get the X-ray luminosity for that SPH particle. Then we obtain for each halo the total X-ray luminosity by summing over all particles within the virial radius. We also obtain the X-ray luminosity weighted temperature and abundances of O and Fe by averaging over these particles. The emission weighting approximates the observational approach in which X-ray spectra are fitted with the APEC or other models to obtain temperature and abundances \citep{dave08}.

\begin{figure}
\centering
\includegraphics[width=0.96\columnwidth]{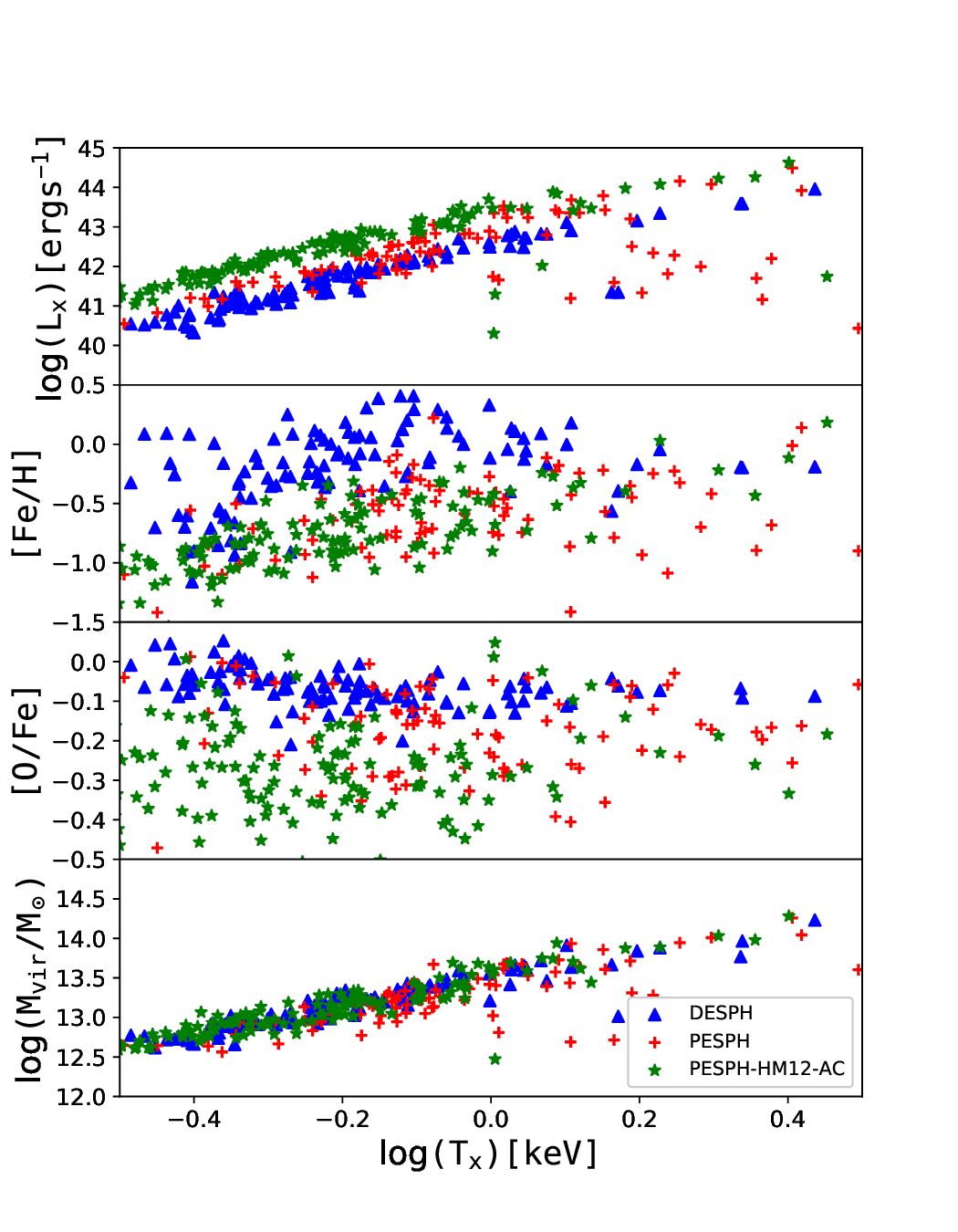}
\caption{The dependence of X-ray luminosity weighted quantities of galaxy groups on their X-ray temperatures. Each point corresponds to a halo from each simulation that contains over 8 resolved galaxies. Haloes from the DESPH, PESPH and PESPH-HM12-AC simulations are plotted as blue triangles, red crosses and green stars, respectively. From the top to the bottom, the total X-ray luminosity $L_x$, iron abundance, alpha enhancement and the $M_{vir}$ of each individual group are plotted against its X-ray weighted temperature $T_x$.
}
\label{fig:xray}
\end{figure}

Fig.\ref{fig:xray} compares how the various X-ray properties depend on the X-ray weighted temperature $T_x$ in our group sample. The bottom panel shows all three simulations agree well on the relation between $M_{vir}$ and $T_x$, but there are large differences in other X-ray properties. The upper panel shows the correlation between the total X-ray luminosity $L_x$ and $T_x$. Gravitational heating alone produces a scaling with $L_x \propto T_x^2$. However, a steeper relation has been observed for clusters and groups \citep{white97, xue00}. The star formation and feedback processes that add non-gravitational energy into the ICM are supposed to alter the relation. This relation, therefore, helps us understand non-gravitational energy sources in our simulations. In this work, a power law scaling is reproduced in every simulation, but the normalisation and scatter vary. In particular, the $L_x$ of the PESPH-HM12-AC groups are systematically higher than the other simulations at fixed $T_x$ by $\sim$ 1 dex. This is likely related to the trend seen in the phase diagram (Fig.\ref{fig:rhot}) that the X-ray emitting gas in the PESPH-HM12-AC simulation is relatively cooler but much denser than in the DESPH simulation. In the PESPH simulation, some massive groups have $L_x$ that is much below the scaling relation. These groups also show much lower enrichment levels than other groups as seen in the next two panels.

\begin{figure}
\centering
\includegraphics[width=0.96\columnwidth]{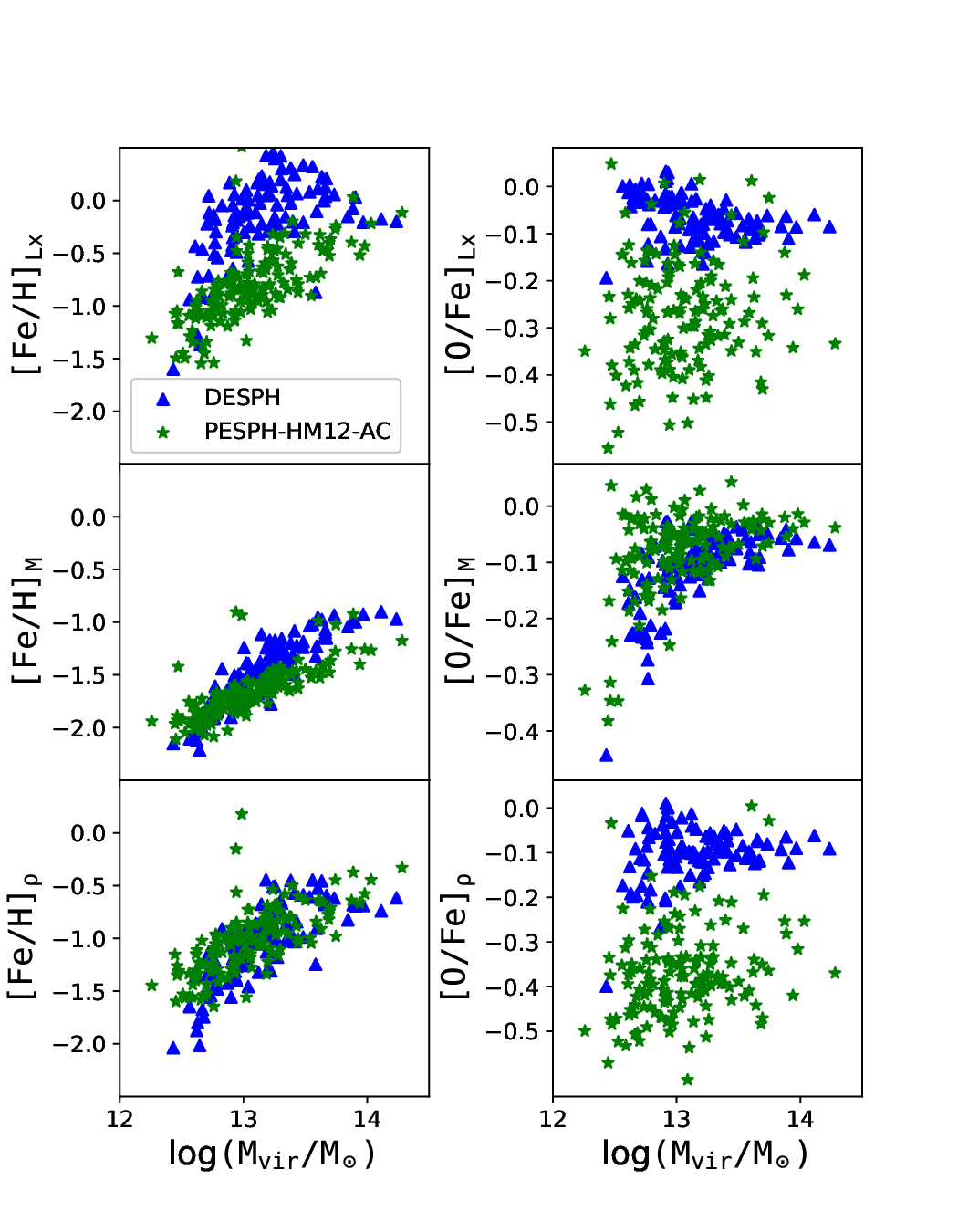}
\caption{Iron abundances [Fe/H] (\textit{left panels}) and $\alpha$ enhancements [O/Fe] (\textit{right panels}) of individual groups as a function of their virial mass $M_{vir}$. The abundances are averaged over all hot gas within the virial radius but are weighted by $L_x$ (\textit{top panels}), mass (\textit{middle panels}) and density (\textit{bottom panels}). In each panel we compare the results from the DESPH simulation (blue triangles) and the PESPH-HM12-AC simulation (green stars).
}
\label{fig:xmetals}
\end{figure}

The second and third panels of Fig.\ref{fig:xray} compare the X-ray luminosity weighted iron abundance [Fe/H] and the $\alpha$ element enhancement [O/Fe] as a function of $T_x$ for selected groups. Both the PESPH and the PESPH-HM12-AC groups have much lower $L_x$ weighted metals than the DESPH simulation and also have much larger scatters in these abundances at fixed $T_x$. To help us understand what causes the differences, we calculate [Fe/H] and [O/Fe] using mass-weighted and density-weighted abundances and compare the results from the DESPH and the PESPH-HM12-AC simulations in Fig.\ref{fig:xmetals}. The mass-weighted [Fe/H] are significantly lower than the flux-weighted values in both simulations, because the flux-weighted abundances are dominated by the densest gas near the group centre, which is also most enriched (see the metal distributions shown in Fig.\ref{fig:metal_rhot_ovi_neviii}). The differences between the flux-weighted abundances seen in the upper panels also disappear in the middle panels after the abundances are re-evaluated and weighted by mass, indicating that the discrepancies mostly owe to the different physical conditions of the densest gas. In the bottom panels, we compare the density-weighted abundances. The distribution of PESPH-HM12-AC groups on this diagram is nearly identical to the upper panels that use flux-weighted abundances. The groups in the DESPH simulation, on the other hand, still have much lower [Fe/H] compared to the flux weighted values, but their [O/Fe] ratios are quite similar now.

In summary, the X-ray properties of galaxy groups change significantly after improving the SPH formulation, artificial viscosity and the timestep regime. These abundances are very sensitive to the enrichment of the hot and dense gas, which is on average lower in simulations that use the PESPH formulation. This result suggests that numerical effects need to be taken into account when making predictions for the X-ray properties of groups and clusters.


\section{SUMMARY} Cosmological baryonic simulations are important for
understanding galaxy formation and evolution. SPH evolves the fluid
equations by treating finite volume fluid elements as particles. The
particle-based nature of SPH allows it to probe a large volume while
maintaining high resolution at the smallest scales, and thus makes
it suitable for simulations in a cosmological context. However,
traditional SPH has difficulties resolving gas dynamics in subsonic
flows, namely it suppresses instabilities and prevents multi-phase
mixing. \citet{hopkins13} posed an alternative form of the SPH equations
of motion that eliminates these problems. We implemented the new formalism
into our GADGET-3 code, along with improved treatments of the artificial
viscosity and the timestep criteria. We performed standard fluid tests
as well as full cosmological tests in a co-moving volume to test the new
algorithm. We also update our metal line cooling model to include non-CIE
effects in the presence of the \citet{haardt12} UV background. The HM12
background is a more up-to-date estimate of the background flux using a
more careful assessment of galaxy and quasar contributions as a function
of time. The \citet{wiersma09} cooling model considers the extra cooling
owing to free electrons from metals, which is potentially significant in
neutral gas.

In this work, we focus on studying the impact of these improvements in
both numerical algorithms and physics to simulations within
a cosmological context. The main conclusion is that the majority of
observables we have studied in previous work \citep{oppenheimer08,
finlator08, oppenheimer10, peeples10a, peeples10b, dave11a, dave11b, oppenheimer12, dave13,
kollmeier14, ford16}, whether on global scales or galactic/halo scales,
are not significantly affected by either the new hydrodynamical models
or the cooling model, while at the same time, the unphysical clumping
of cold gas owing to the deficiencies of the original algorithms is
effectively removed. However, observables related to the amount or metallicity of hot gas are affected.

We do not extensively explore how the differences caused by the different SPH schemes will be affected by numerical resolution, but the effect of a modest increase in resolution such as in a zoom-in simulation is likely to be unimportant. The standard hydrodynamic tests, in which the differences caused by different numerical algorithms are distinct, have much higher resolution than even the zoom-in simulations. The major difference between the DESPH and the PESPH is the efficiency of phase mixing, which is intrinsic to the SPH formulation and independent of resolution. \citet{fire2} discuss the effects of SPH implementations in the zoom-in, Feedback in Realistic Environments (FIRE) simulations. They find that many of their qualitative conclusions are robust to SPH implementations, but the efficiency of phase mixing has important effects on CGM properties and subsequent cooling and re-accretion of the CGM gas in massive, hot halos. This general conclusion is consistent with our findings from lower resolution simulations on cosmological scales.

The cosmological simulations in this work do not include AGN feedback, nor did our previous simulations. Therefore, we do not assess the impact of different SPH schemes on AGN feedback in this work. \citet{schaller15} demonstrate that the pressure-entropy SPH enhances phase mixing between accreted gas and the AGN driven bubble and, therefore, affects the efficiency of AGN quenching. This leads to changes in the stellar mass content and the X-ray emission from the intragroup medium in massive halos. However, this effect of changing numerical schemes likely depends on the resolution and could be mitigated with a recalibration of the subgrid parameters that govern the efficiency of the AGN feedback. The nIFTy comparison project also demonstrates that a variety of simulations, which employ a variety of different numerical and feedback models, produce similar gas properties for a galaxy cluster \citep{sembolini16b}, even though the different numerical schemes without the subgrid models lead to significant differences in the gas profiles of the simulated clusters \citep{sembolini16a}. This further indicates the importance of subgrid models and their calibration.

We conclude:

\begin{enumerate}
  \item The PESPH formulation alone significantly enhances hot mode
  accretion in massive haloes at all redshifts, indicating that cooling
  is more efficient in these systems, and that some cold mode accretion is transformed into hot mode accretion owing to better resolved shocks. In the simulations that employ
  momentum-driven wind feedback, cold mode accretion and wind reaccretion
  still dominates at early and late times, respectively, regardless
  of the numerical scheme used. The total amount of accretion only
  slightly increases owing to more efficient wind reaccretion at lower
  redshifts. The impact of the PESPH formulation is much stronger in the \textit{nw} simulations where wind-reaccretion is not present. The hot accretion
  fraction increases substantially at most redshifts and especially at
  high redshifts, where it becomes dominant in $\log(M_h/M_\odot) > 11$ haloes. Consistent with our previous work, cold mode accretion dominates the overall growth of the galaxy population at high redshift and the growth in low mass haloes at low redshift, while wind re-accretion dominates galaxy growth in massive haloes at $z < 2$.

  \citet{nelson13} investigates the gas accretion problem
  using a set of non-feedback cosmological simulations simulated
  with the moving mesh code AREPO. Using a similar temperature cut
  to distinguish cold and hot mode accretion, they find the cold
  mode accretion is nearly completely removed from $z=2$ haloes with
  $\log(M_h/M_\odot) \sim 12$. In our simulations, however, despite
  the fact that the cold accretion fraction significantly drops owing
  to the PESPH formulation, it is still present at the $\sim 20\%$
  level in most haloes at $z=2$. This disagreement is not likely caused
  by the artificial clumps seen in DESPH simulations, since the PESPH
  formulation has eliminated these features and yet is still able to
  fuel central galaxies of the massive haloes with cold gas. However,
  a detailed comparison with their work is not feasible owing to both
  the different sub-grid assumptions and the accretion tracking method.
  \item The PESPH formulation, along with the \citet{cd10} viscosity and
  the \citet{durier12} timestep criteria enhances the star formation rate
  in our fiducial \textit{ezw} model by as much as 20\% after the star
  formation peak at $z\sim 3$, increasing the number density of massive
  galaxies at low redshifts. The non-CIE cooling model suppresses cooling
  and star formation by a comparable amount after the star formation
  epoch, resulting in a SFH close to the original one. The enhanced star
  formation in massive systems owing to numerical improvements, however,
  is much smaller than the discrepancies between the simulations and
  observations. The requirement of additional feedback, either AGN or a
  different supernova feedback algorithm (Huang et al in prep.) that quenches
  star formation in massive galaxies is not alleviated. Moreover,
  as pointed out by \citet{hu14}, the impacts of PESPH and artificial
  viscosity on star formation and feedback probably depends on the
  feedback implementation. In the thermal feedback scheme, the mass
  loading is significantly reduced by the enhanced mixing between SNe
  heated gas and the surrounding cool gas, while our kinetic feedback is
  much less sensitive to the efficiency of mixing. 

  \item Using
  the fiducial PESPH-HM12-AC simulation that employs the state-of-art
  numerical algorithms, we confirm the ``photon underproduction crisis'' reported in \citet{kollmeier14}. Namely, our simulation strongly overproduces
  the column density distributions of HI absorbers at low redshift if we adopt a uniform HM12 ionisation
  background. We find that the HI column density distribution
  traced by Ly$\alpha$ absorption is very robust to the choices of
  hydrodynamics. However, changing from the HM01 background to the HM12
  background boosts the amount of HI absorbers by a factor of 3, which is
  consistent with the findings from \citet{kollmeier14}. The conclusion
  that the predicted Ly$\alpha$ forest is robust to hydrodynamics is also
  reported in \citet{gurvich16}, who shows that the differences between
  a traditional SPH (DESPH) code and the moving mesh code are small in
  the low column density range studied here.  \item The distributions
  of baryons in the density-temperature phase diagram are quite similar
  in general, with the exception that the PESPH formulation leads to
  more hot gas at higher densities, and the non-CIE metal cooling model
  results in a larger scatter in the condensed gas phase. The mass
  fractions of baryons that reside in the various phases: diffuse, WHIM, hot,
  condensed, and stars, are surprisingly similar in all simulations at
  most redshifts. The hot gas fraction in galaxy groups is enhanced by a
  small amount in the wind models compared to the ones without galactic
  winds, indicating that the contribution of galactic winds to the hot intragroup medium is small. In spite of the different feedback models, the amount of
  hot gas is quite robust to the numerical schemes.

  \item The PESPH formulation enhances metallicity in the WHIM and hot halo gas. As a result, the PESPH simulations generally produces more absorbers
  for some of the high ions, e.g., CIV, OVI, NeVIII, SiIV,
  on both global scales indicated from their column density distributions
  and within $\log(M_h/M_\odot)=12$ haloes at low redshift. The non-CIE
  cooling and the HM12 background reduce the amount of weak absorption,
  except for NeVIII absorption, which is even further enhanced by these changes at all
  column densities.  

  
  \item The X-ray group properties change significantly
  with the improvements in the hydrodynamics, but they are insensitive to the
  adopted cooling model. The X-ray luminosity weighted iron abundance
  and $\alpha$ enhancement in the group sample are both greatly reduced
  and have much larger scatter in the PESPH and PESPH-HM12 simulations
  compared to original results from the DESPH simulation. These differences mostly owe to the different metallicities in the densest gas near the group centre in these simulations. There
  are also more outliers from the $L_x-T_x$ scaling relation, which correspond
  to groups that have the lowest X-ray weighted metallicity. A more
  careful analysis of X-ray properties is left for future work.

  \item In Appendices C, D and E, we show that the change from our previous DESPH formulation to the new PESPH-HM12-AC prescription has little impact on many of the key characteristics of galaxy evolution: star formation histories as a function of halo mass, the HI mass function, and the redshift-dependent scaling relations between galaxy stellar mass and specific star formation rate, cold gas fraction, and gas phase metallicity. These characteristics and many of the others mentioned above can, therefore, be used to set constraints on the physics of stellar and AGN feedback, galactic outflows, and metal mixing in the CGM, without high sensitivity to numerical or microphysical uncertainties.
\end{enumerate}

In summary, we test the PESPH formulation, along with other improved
numerical technologies and physical models, in full cosmological
hydrodynamic simulations that are implemented with state-of-art models
for baryonic physics. The new implementations successfully removed
long standing numerical artefacts without significantly affecting most
predictions from previous simulations.

\section*{Acknowledgements}

We thank the anonymous referee for useful comments. We thank Amanda Ford for sharing her analysis code for generating mock QSO lines, and Volker Springel
for providing the GADGET-3 code. We have used SPLASH \citep{price07}
for visualisation. We acknowledge support by NSF grant AST-1517503,
NASA ATP grant 80NSSC18K1016, and HST Theory grant HST-AR-14299.
DW acknowledges support of NSF grant AST-1516997.

\bibliography{references}

\appendix

\section{The Sedov-Taylor Blastwave Test} 
The Sedov-Taylor blast wave test starts with a point
explosion within a uniform medium that creates a strong shock wave,
which sweeps through its surroundings. The numerical solution to this
problem is very sensitive to the shock capturing capabilities owing to
the sharp entropy discontinuity at the shock front. In SPH, shocks are
captured by numerical dissipation through artificial viscosity. Therefore,
this problem is a strong test of the artificial viscosity scheme and the
viscosity parameters. We set up a three dimensional uniform lattice in a
$6^3\ \mathrm{kpc}^3$ cube with 64 particles on a side. The volume has a
uniform initial density of $1.24 \times 10^7 M_\odot\ \mathrm{kpc}^{-3}$
and temperature of $10\ K$. We start the simulation by injecting $E =
6.78 \times 10^{53}$ ergs of energy into the eight central particles,
resulting in a initial entropy contrast of $3\times10^6$ and a Mach
number of $\sim 1000$.

\begin{figure} 
\centering
\includegraphics[width=1.00\columnwidth]{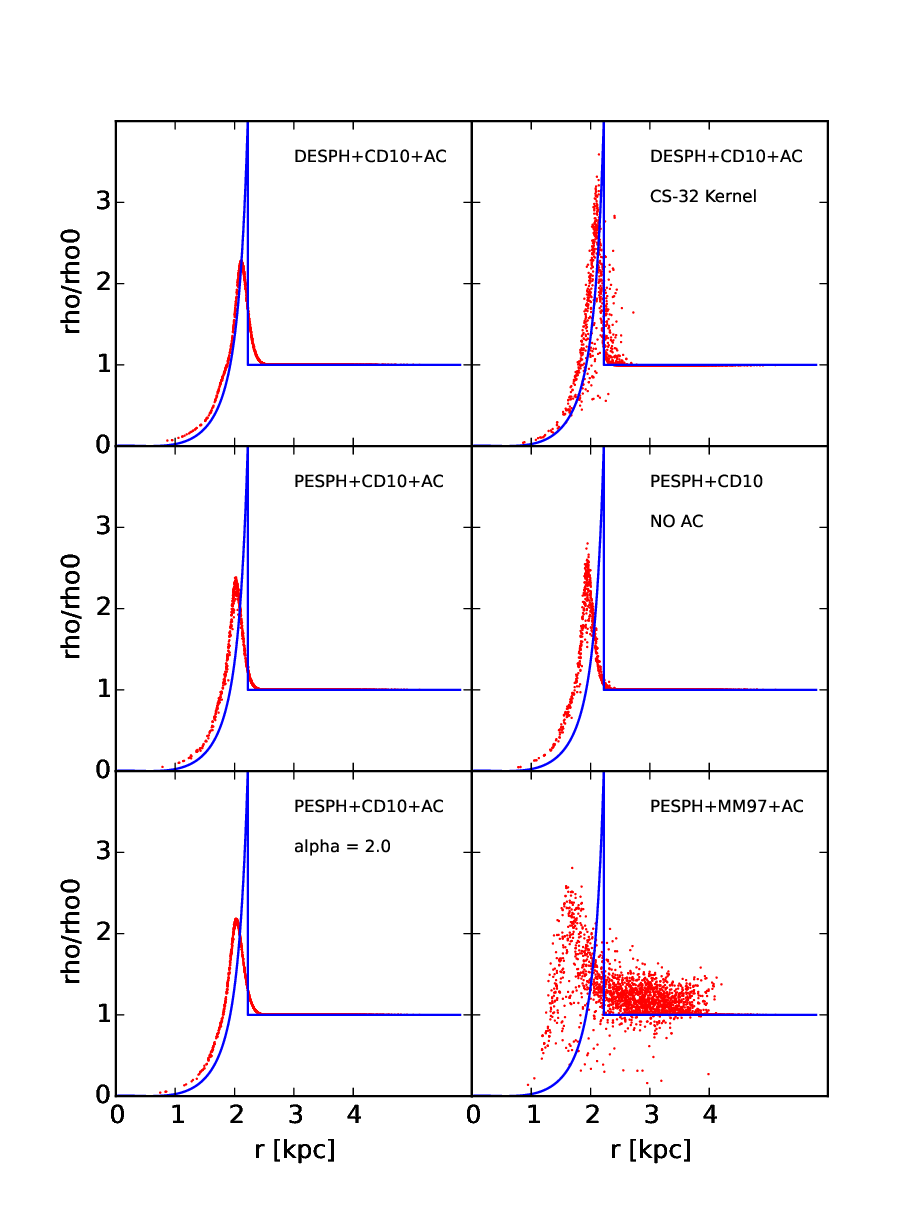}
\caption{Numerical solutions of the Sedov-Taylor blastwave tests
are compared to the analytic solution. The analytic solution for the
overdensity as a function of radius from the energy injection point is
shown in blue. In each panel, each red dot represents an SPH particle from
the simulation. One out of a thousand particles are displayed here. The
first row uses the DESPH formulation in addition to the \citet{cd10}
viscosity and artificial conduction with our fiducial parameters. The
right panel of the first row uses the old cubic spline kernel instead of
our fiducial quintic kernel. The other four panels use the PESPH formulation
with different choices of viscosity and conduction: In the second row,
the left panel is the fiducial numerical choice for our cosmological
simulations; the right panel does not have the artificial conduction. In
the third row, the left panel uses $\alpha_{max}=2.0$ for the viscosity,
larger than our fiducial $\alpha_{max}=1.5$ value; the right panel uses
the \citet{mm97} viscosity with the $\alpha_i$ starting from the minimum
value $\alpha_{min}=0.2$.  } \label{fig:sedov} 
\end{figure}

The Figure \ref{fig:sedov} shows the results from six
simulations. We find that to reach a good agreement between the
numerical result and the analytical solution for the Sedov-Taylor test,
a higher order kernel with a large enough number of neighbours, combined
with a timestep limiter and a sensitive artificial viscosity switch
are crucial. Simulations that incorporate these features reproduce the
analytic solution reasonably well. The shock fronts align with each other,
although the numerical solutions have a reduced peak density and some
post-shock ringing effects.

The PESPH formulation performs worse than the DESPH formulation, because
the entropy-weighting nature of PESPH exaggerates noise at the entropy
jump, which is very sharp in this test, as is also found in other work
\citep{hu14, read12}. Also, the shock fronts in PESPH simulations are
delayed compared to the analytic solution, a phenomenon also shown in
\citet{read12} with their RT formulation, which similarly evaluates the
density using entropy weighting.

The \citet{mm97} viscosity (bottom right panel) is very poor at shock
capturing. Note that in this test, the $\alpha$ parameters of the
particles start from the minimum value as in the cosmological simulations
where the viscosity is applied. The density profile is very noisy at
the shock front and particle penetration is evident. Those particles
in the shock are not sufficiently shocked owing to the slowly-growing
$\alpha$. Since the energy injection is not isotropic, particles that
lie in the direction of the energy injection propagate much further than
the surrounding particles. The \citet{cd10} viscosity, on the other hand,
prepares these particles by boosting the $\alpha$ parameter to the maximum
value even before the shock front passes and thus effectively captures the
shock. Our fiducial choice of $\alpha_{max}=1.5$ is a compromise between
efficient shock capturing and lowering unwanted dissipation. The bottom
left panel shows the results using a slightly larger $\alpha_{max}=2.0$,
but the improvement is only marginal.

The choice of kernel is not trivial in this test. A low order cubic
spline kernel with 32 neighbours, even in an idealised case with DESPH,
\citet{cd10} viscosity, and artificial conduction, still produces a lot
of noise at the shock front (top right panel). This is also similar to
\citet{read12}, where they found a cubic spline kernel with 42 neighbours
results in a much more noisy shock front than a higher order HOCT kernel
with 442 neighbours, despite the fact that the HOCT kernel lowers the
spatial resolution by a factor of 1.5.

Artificial conduction reduces the noise at the shock front by smoothing
the entropy contrast. It can be effective in reducing noise within the
PESPH formulation, but the effect is not significant in our tests. Note
that our artificial conduction coefficient is only one quarter of the
commonly adopted value.

Finally, we need to emphasise that all the simulations above 
use the timestep limiter that adjusts the timesteps of particles
in the vicinity of large energy fluctuations, so that these particles
are able to react to the approaching shock in time. We use a Courant
factor of 0.05 to determine the timesteps in general. However, without
the timestep limiter, the results degrade significantly.  

\section{Artificial Conduction in a Cosmological Volume}
Artificial conduction is introduced as a pure numerical
tool whose purpose is to smooth out fluid discontinuities in some
high resolution standard test problems and thus to achieve realistic
solutions to these problems. However, in a large cosmological volume,
such high resolution is unattainable. Moreover, artificial conduction
could lead to spurious cooling near galaxies and compromise the validity
of the results. Since the radiative cooling rate depends critically on
the gas temperature especially at the disc-halo interface, non-linear
effects could magnify the effects of conduction on gas cooling. In this
section, we investigate the role of artificial conductive cooling in
our cosmological tests and compare it to the amount of radiative cooling.

\begin{figure} 
\centering
\includegraphics[width=1.00\columnwidth]{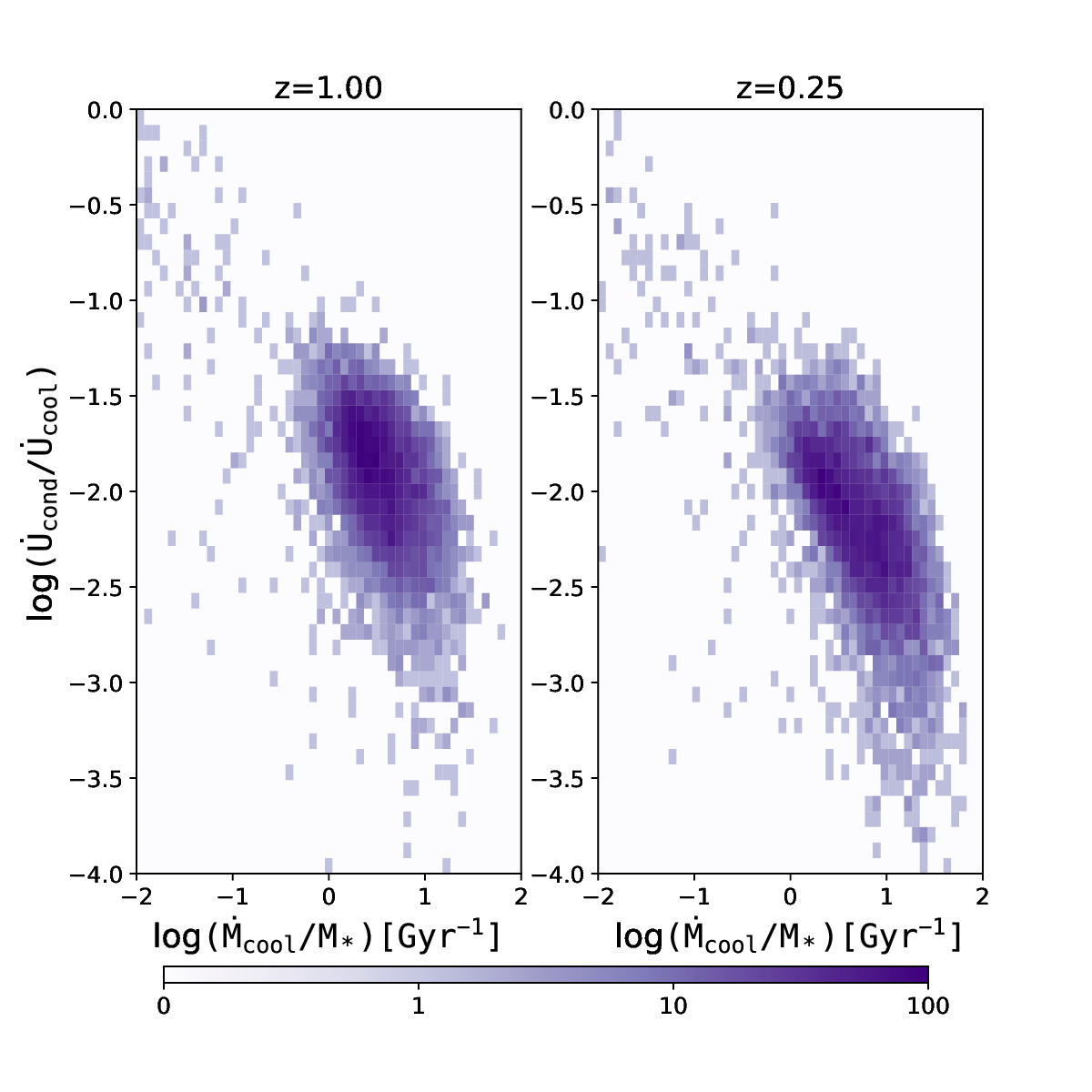}
\caption{The cooling ratios between artificial conductive cooling
and net cooling (including conductive cooling and heating sources) are plotted against the specific net cooling rate for each halo
from the PESPH-HM12-AC simulation. Results at $z=1$ and $z=0.25$
are shown in the left and right panels, respectively. Here radiative
cooling is computed only from non-SF particles. Note that the net
cooling rate $\dot{U}_{cool}$ includes all cooling and heating sources,
so the ratio can be large where cooling balances heating so that the net cooling rate is small. In GADGET-3,
it is not straightforward to separate the radiative cooling rate from the
net cooling rate, so it is hard to directly compare conductive cooling
to radiative cooling.  } \label{fig:duratio} 
\end{figure}

We take snapshots at $z=1$ and $z=0.25$ from our fiducial PESPH-HM12-AC
simulation and measure the cooling rates in all the SO haloes identified
in each snapshot. Figure \ref{fig:duratio} compares the ratio between the
total conductive cooling rate $\dot{U}_{cond} = \sum_im_i\dot{u}_{cond,i}$ and the
total net cooling rate $\dot{U}_{cool} = \sum_im_i\dot{u}_{cool,i}$ in each halo, as a function of the specific cooling rate,
defined as $\lambda_{cool} \equiv \dot{M}_{cool}/M_{*}$, where $M_{*}$
is the stellar mass of the central galaxy, and $\dot{M}_{cool} \equiv \sum_im_i\dot{u}_{cool,i}/u_i$.

Conductive cooling is almost always small and sub-dominant
compared to the net cooling. In a few haloes, where the ratio
$\dot{U}_{cond}/\dot{U}_{cool}$ approaches unity, the specific
cooling rates are too small to be important to the thermal histories
of these haloes. In haloes where conductive cooling is fast, the net cooling is orders
of magnitude faster, making the contribution from artificial conduction
negligible.

There are elements in our conduction scheme that attempt to restrain
conduction to only fluid discontinuities. Our requirement that
conduction occurs only when the signal velocity between particles,
$v_{sig}=c_i+c_j-3\omega_{ij}$, is positive suppresses conduction in
non-converging flows. Also, the factor $L_{ij}$ reduces conduction
in pressure equilibrium. The artificial conduction coefficient that
we use ($\alpha_{cond} = 0.25$) is only one quarter of the commonly
assumed value, but increasing conduction by a factor of 4 does not
significantly alter the results above. We did not specifically study
the heating from conduction. The artificial conduction scheme that
we adopt is conservative so that the amount of conductive heating is
equal to the conductive cooling. The cool gas will thus quickly dissipate
the heat gained from conduction and return to thermal equilibrium on a
very small time-scale. Therefore, we are confident from these tests that
the spurious numerical dissipation from our implementation of artificial
conduction has a minimal effect on the growth and evolution of galaxies
in a cosmological volume.

\section{Star Formation Histories} In previous sections we
have shown that the accretion rate, total stellar content, and the stellar
mass function are barely affected by the improved numerics. Here we examine the star formation histories (SFH) from our simulations
and how they are affected by the numerics. The SFH reflects the overall
efficiency of galaxy formation and relies on critical processes such as
accretion efficiency, cooling, and feedback.

\begin{figure} 
\centering
\includegraphics[width=0.96\columnwidth]{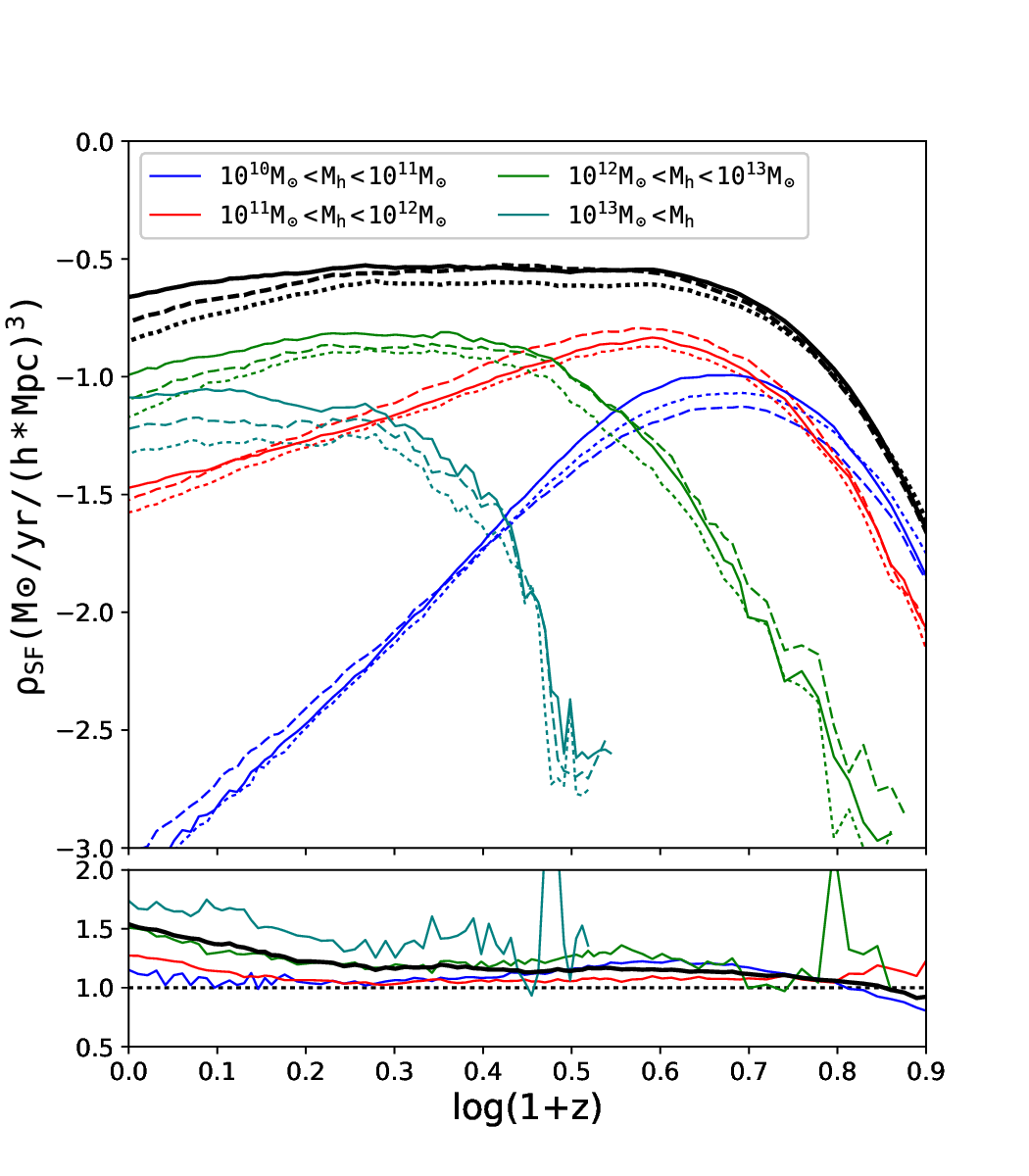} \caption{The
upper panel shows the star formation histories of three simulations with galactic winds (\textit{ezw}). The
dotted, dashed and solid lines correspond to the DESPH,
PESPH, and PESPH-HM12-AC, respectively. The black lines show the global
star formation density as a function of redshift. The blue, red, green
and teal lines show the summed star formation density at any time from
individual haloes grouped according to their total mass $M_h$ at that
redshift as labelled. The bottom panel shows the ratios of the SFRDs
from the PESPH-HM12-AC and the DESPH simulations for each halo mass bin,
using the same colour scheme.  } \label{fig:sfh} 
\end{figure}

Figure \ref{fig:sfh} presents the global SFH of the DESPH, PESPH, and
PESPH-HM12-AC simulations. We plot the contribution to the star formation
rate from haloes that are grouped by their total mass. At each redshift,
the SFR of each ISM particle is computed based on the subgrid SF model,
then the SFR of each halo is summed over all the ISM particles that it
contains. We have included contributions from sub-resolution galaxies
since we are focusing here on the total amount of baryons in stars and
the effect of including these galaxies is actually small.

The overall evolution of the star formation histories is similar
in the three simulations. All SFHs are characterised by a sharp peak
at $z\sim3$ and a gradual decline afterwards. The total SFHs in the
DESPH and PESPH runs agree quite well before the star formation epoch,
then the PESPH starts to form $\sim 20\%$ more stars than the DESPH
all the way down to $z=0$. This offset is caused only by the different
treatment of hydrodynamics. Massive haloes with $\log(M_h/M_\odot) > 11$
in the PESPH simulation form stars at a higher rate over the entire time,
causing the overall $20\%$ more stars after $z\sim3$.

Compared to the PESPH results, the new cooling model and background
in the PESPH-HM12-AC simulation suppresses star formation in all
haloes after the star formation peak, but has little effect at higher
redshift. In the \citet{wiersma09} cooling model, species that dominate
cooling in warm gas are over ionised relative to collisional equilibrium in the presence of the photoionising
background, shifting their cooling curves toward lower
temperature, suppressing the net metal-line cooling efficiency and
the star formation rate. Therefore, the differences between PESPH and
PESPH-HM12-AC are primarily caused by the cooling models used.

Coincidentally, our implementation of hydrodynamics and physics works
against each other in regulating the star formation processes. The
combined effects produce a star formation history that closely resembles
the results from our previous model, except for a short period of slightly
enhanced star formation between $z=2$ and $z=3$. It is clear from the
bottom panel that the massive haloes at these redshifts contribute most
to the overall enhanced star formation.

\section{HI Mass Function}
\begin{figure}
  \begin{center}
    \includegraphics[width=0.97\columnwidth]{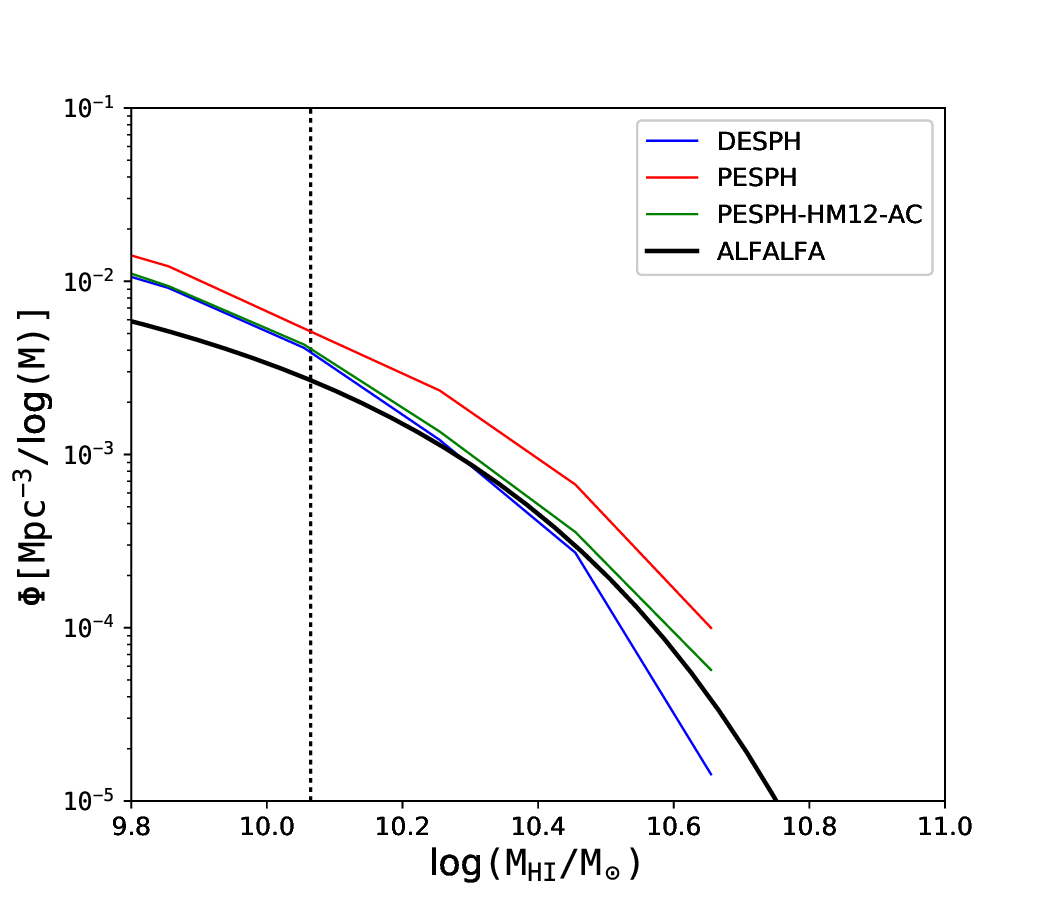}
    \caption{The
neutral hydrogen mass functions (HIMFs) at $z=0$. Blue, red and
green lines represent results from the DESPH, PESPH and PESPH-HM12-AC
simulations, respectively. We also include the Schechter fit of the
HIMF from the $\alpha.40$ sample of ALFALFA survey \citep{haynes11} as
thick black solid line. The vertical dotted line indicates the stellar
mass resolution limit, which is comparable to the neutral hydrogen mass
limit \citep{dave13}.  } \label{fig:himf} \end{center} \end{figure}

Figure \ref{fig:himf} compares the neutral hydrogen mass functions
(HIMFs) from our simulations. We calculate the neutral hydrogen fraction
of every gas particle following \citet{dave13}. In summary, we first
calculate the optically thin limit of the neutral fraction by solving
for ionisation equilibrium based on local density, temperature and the
radiation field. Then the column density is integrated for each gas
particle, assuming a uniform distribution of neutral hydrogen within
a sphere with a radius equal to its smoothing length. If the column
density of HI calculated as such exceeds a threshold of $N_{HI,lim}
= 10^{17.2}\ \mathrm{cm^{-2}}$ where the gas becomes optically thick,
we compute the HI content by including the self-shielding of
neutral hydrogen, and compute the fraction of the molecular hydrogen
component by using the observed pressure relation \citep{leroy08}. The
HI mass of each galaxy is summed over all gas particles within a sphere
that extends to the outermost radius of the galaxy.

The three simulations generally agree with each other. The galaxies from
the PESPH simulation show slightly more neutral gas content than those from the DESPH simulation. Two comparable effects contribute to this
small yet non-negligible discrepancy. First, the galaxies from the PESPH
simulation assemble in slightly denser regions where the neutral fraction
is higher. Second, the spherical region surrounding PESPH galaxies
encloses more gas particles. Ram pressure stripping on resolved scales,
which is supposed to be enhanced by PESPH, does not significantly lower
the neutral gas content in satellite galaxies.  Adding the improved physics
decreases the HI masses almost back to their original DESPH values.

\section{sSFR, MZR and Gas Fraction}
Observed galaxy populations often present tight scalings between
various properties such as the star formation rate, metallicity, cold
gas fraction and the stellar mass. These scaling relations result from
a complicated interplay between physical processes that are critical to
galaxy formation and evolution, and reproducing them has been a key test of cosmological
simulations. In this section, we demonstrate that these properties of
simulated galaxies are also robust to the different numerical models
and physics explored in this work.

\begin{figure*}
\centering 
\includegraphics[width=1.80\columnwidth]{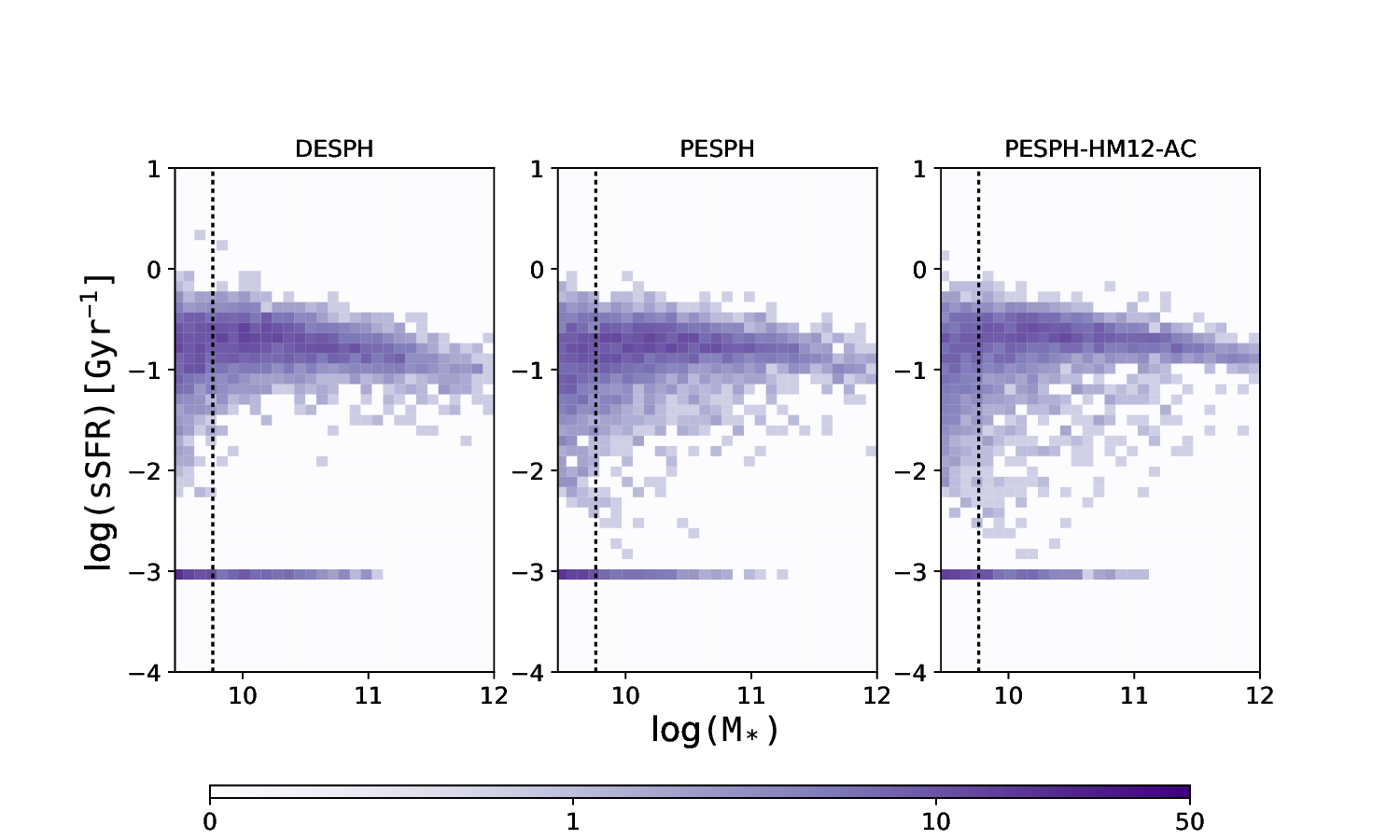}
\caption{The specific star formation rates (sSFR) of simulated galaxies as
a function of their stellar masses at $z=0$. The horizontally stretched
region around sSFR $= 10^{-3}\ \mathrm{Gyr}^{-1}$ is populated by
galaxies that contain no gas (and hence no star formation), so we assign
an arbitrary sSFR. The dotted vertical line indicates half the resolution limit
(32 gas particles). The number density of galaxies in each bin is colour
coded in purple.  } \label{fig:color_mag} 
\end{figure*}

Figure \ref{fig:color_mag} compares the specific star formation rate as
a function of stellar mass for individual galaxies from the simulations
at $z=0$. We calculate the sSFR by averaging the star formation rates
of all SPH particles with densities above the star forming threshold in
each galaxy. There are also quiescent galaxies in our simulations that
only contain star particles and no gas particles. We assign an arbitrary
value of $sSFR = 10^{-3}\ \mathrm{Gyr}^{-1}$ to these galaxies. The two PESPH simulations produce slightly more low sSFR
galaxies, but the overall results of all these simulations are very similar. Cosmological simulations without
feedback do not reproduce the observed red population of local galaxies
\citep{keres09b}, and the kinetic feedback model employed here with
improved numerical schemes is still not able to resolve this discrepancy. This finding strengthens the conclusion that a new feedback scheme, possibly adding AGN feedback, is needed to explain the observed population of passive galaxies.

\begin{figure*}
\centering
\includegraphics[width=1.20\columnwidth,angle=0]{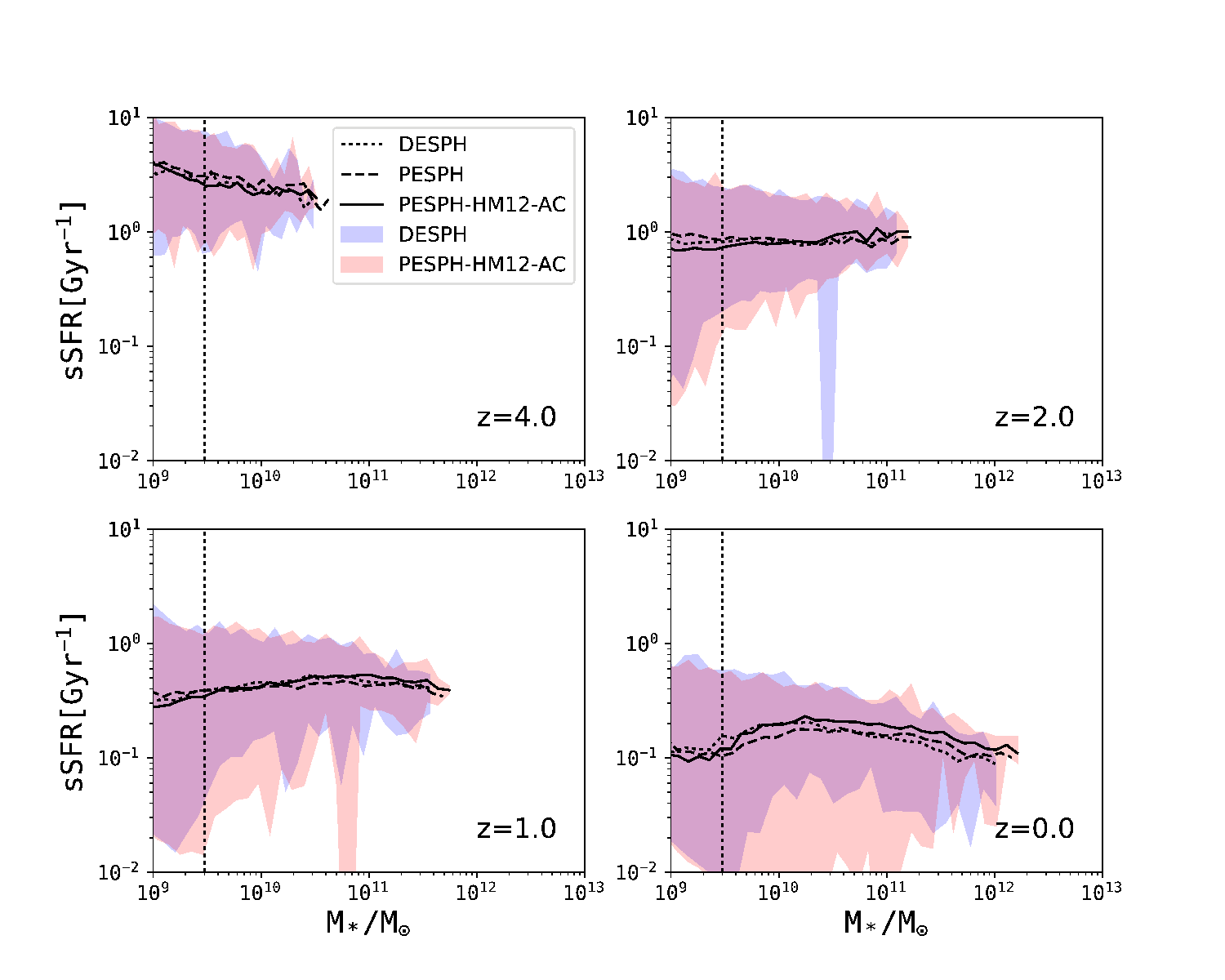}
\caption{The relation between sSFR and stellar mass at four different
redshifts. The dotted, dashed and solid lines are medians of the sample
from the DESPH, PESPH and PESPH-HM12-AC simulations, respectively. The
shaded area contains 95\% of galaxies in each mass bin (representing only
galaxies from DESPH and PESPH-HM12-AC). The dotted vertical line indicates
half the resolution limit (32 gas particles).  } \label{fig:ssfr}
\end{figure*}

\begin{figure*} 
\centering
\includegraphics[width=1.20\columnwidth,angle=0,trim=0 0 0 0.5,clip=true]{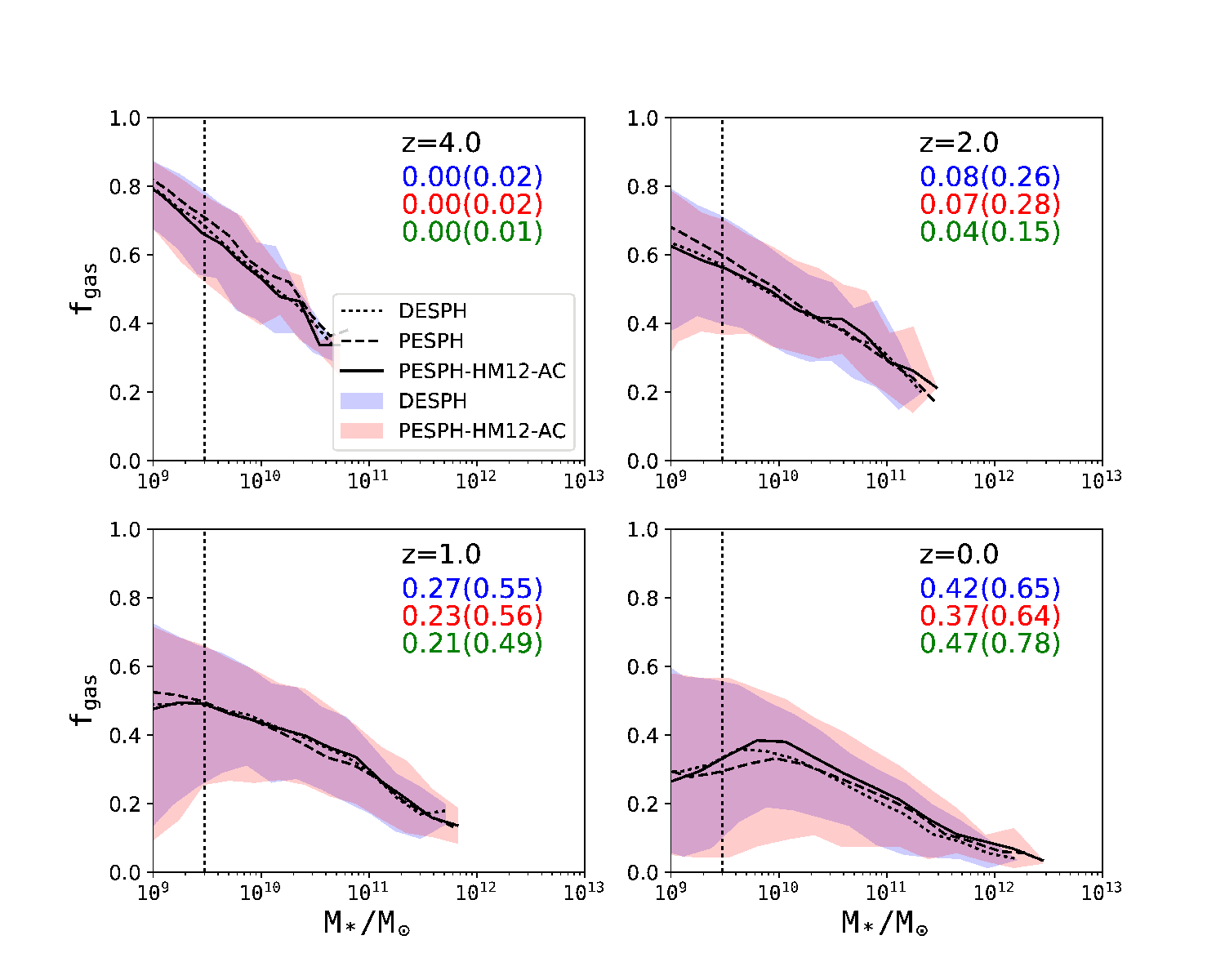}
\caption{The relation between gas fraction and stellar mass at
four redshifts. The dotted, dashed and solid lines are the medians
of the sample from DESPH, PESPH and PESPH-HM12-AC simulations,
respectively. Black lines include all galaxies while green lines
only include satellite galaxies. The shaded area contains 95\% of
galaxies in each mass bin, including only the galaxies from the DESPH
and PESPH. The dotted vertical line indicates half of the resolution
limit (32 gas particles). The floating numbers in each panel indicate
the fraction of gas-free galaxies in each simulation, with the number
within the parentheses indicating the gas-free satellite galaxies.  }
\label{fig:fgas} 
\end{figure*}

The sSFR-$M_*$ relation also agrees very well between simulations at
most redshifts (Figure \ref{fig:ssfr}). The shaded areas encloses 95\%
of the galaxies from each sample. Here we only compare the scatter of
the relation from the DESPH and the PESPH-HM12-AC simulation to highlight
the differences between our old and new fiducial runs. The scatter also
agrees quite well between the old and new model at all redshifts, except
that the new fiducial simulation PESPH-HM12-AC contains some galaxies
with very low sSFR at $z=0$ and $z=1$.

The gas content within galaxies is balanced by accretion and gas
consumption owing to star formation. Figure \ref{fig:fgas} shows the
trend of gas fraction within galaxies with their stellar mass. The
gas fraction $f_{gas}$ is defined to be the mass of all gas particles
within a galaxy divided by the total mass. The galaxies that are not star
forming have been excluded from this sample. The same trend that the gas
content is on average lower in more massive galaxies is observed in all
simulations. Moreover, both the medians and the distributions of the gas
fractions at any particular stellar mass bin are very similar among the
simulations. The regulation of the gas reservoir in individual galaxies
is mostly affected by the star formation and feedback prescriptions,
while the choice of hydrodynamical method and background appears
to have little impact. Apart from the general agreement, the PESPH
simulation has slightly more gas-free galaxies, but most of them are
under resolved. These galaxies are not counted while drawing the median,
but the effect of including them would be small. In addition, galaxies
in the PESPH simulations have a slightly larger gas fraction in general
than DESPH galaxies at all redshifts, as indicated both by the median and
the shaded area. It suggests that the gas supply in PESPH galaxies must be
generally more efficient at refilling their consumed gas to compensate for
the higher star formation and outflow rate. This is consistent with our
previous finding that PESPH galaxies have a higher overall accretion rate.

\begin{figure*}
\centering
\includegraphics[width=1.20\columnwidth,angle=0,trim=0 0 0 0.5,clip=true]{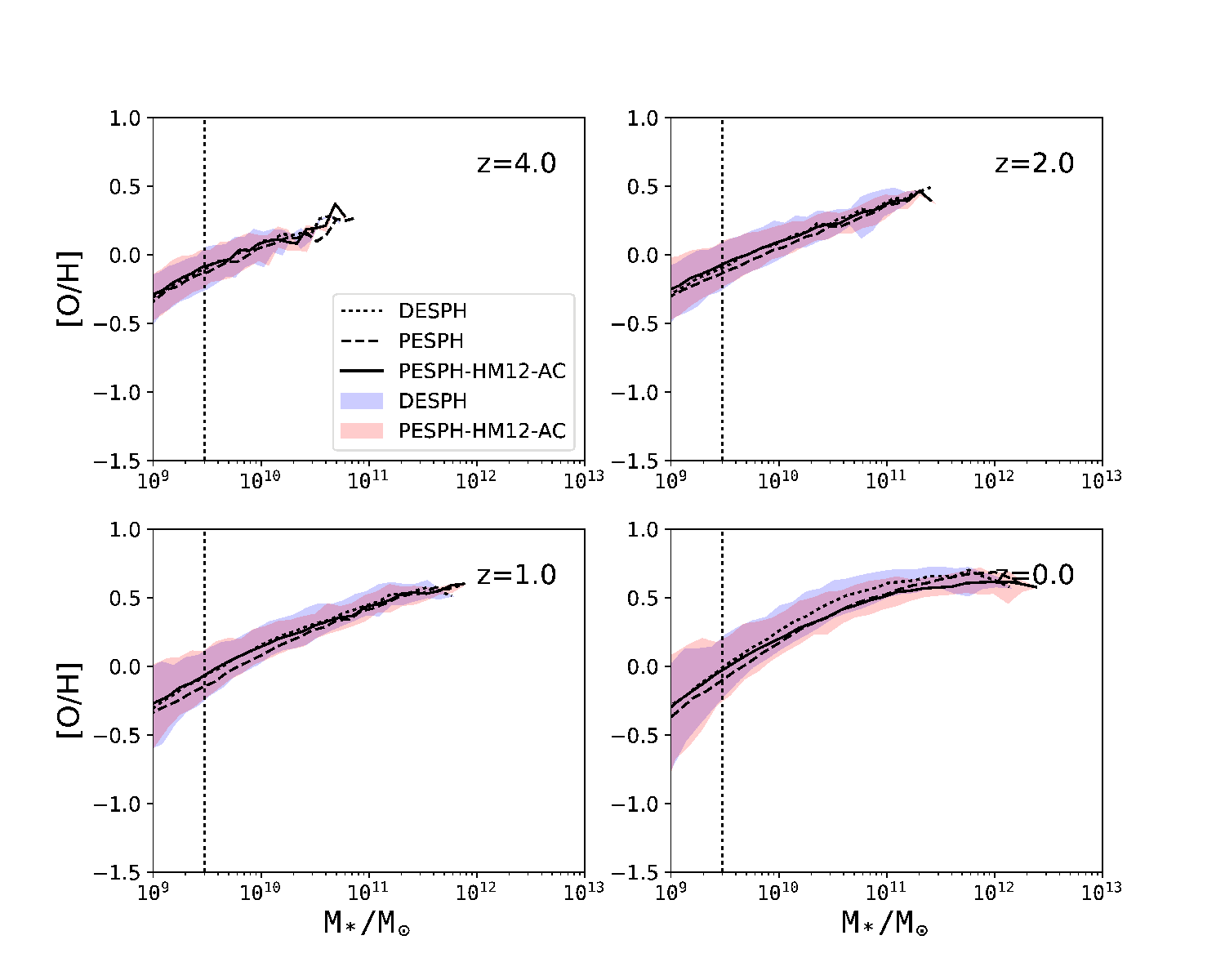}
\caption{The stellar mass-gas metallicity relation at four redshifts. The
metallicity is weighted by the star formation rate of all gas particles
within each galaxy. The dotted, dashed and solid lines are the running
medians of the sample for DESPH, PESPH and PESPH-HM12. The shaded
area contains 95\% of the galaxies in each mass bin, including only
the galaxies from the DESPH and PESPH-HM12 simulations. The dotted
vertical line indicates half the resolution limit (32 gas particles).
} \label{fig:mzr} 
\end{figure*}

Figure \ref{fig:mzr} compares the mass-metallicity relation (MZR) at
four redshifts. Our version of GADGET-3 traces the abundances of C, O,
Si and Fe separately. Here we estimate the metallicity from the oxygen
abundance normalised to the solar value, assuming a solar oxygen mass fraction of
0.009618 \citep{anders89}. The metallicity of each galaxy is averaged
over all the ISM particles weighted by their star formation rate. This
definition avoids a bias to gas particles lying in the outskirts of
galaxies and mimics the observational approach of measuring metallicity
\citep{dave11b}. The slope and scatter of the relation are both
preserved by the PESPH and PESPH-HM12 simulations, while the metal
content in both simulations decrease at all mass scales, most clearly
seen at $z=0$, where the intergalactic gas metallicity generally drops
by 0.1 dex. The new hydrodynamics that boosts star formation in massive
galaxies facilitates metal enrichment as well, maintaining the shape
of the relation, while the altered balance between inflow and outflow
could have lowered the metal content.


\bsp

\label{lastpage}

\end{document}